\documentclass{aa}  

\usepackage{graphicx}
\usepackage{txfonts}
\usepackage{mathtools}
\usepackage{wasysym}
\usepackage{xcolor}

\begin{document}

   \title{A critical assessment of the applicability of the energy-limited approximation for estimating exoplanetary mass-loss rates}


   \titlerunning{A critical assessment of the applicability of the energy-limited approximation}
   \authorrunning{Krenn et al.}

   \author{A. F. Krenn\inst{1,2}
          \and
          L. Fossati\inst{2}
          \and
          D. Kubyshkina\inst{3}
          \and
          H. Lammer\inst{2}
          }

   \institute{University of Technology Graz, Rechbauerstraße 12, A-8010 Graz\\
              \email{a\_krenn@gmx.at}
         \and
             Space Research Institute, Austrian Academy of Sciences, Schmiedlstraße 6, 8042 Graz
        \and
            School of Physics, Trinity College Dublin, the University of Dublin, College Green, Dublin-2, Ireland
            }

   \date{Received: 28.01.2021 Accepted: 20.04.2021}

 
  \abstract
   {The energy-limited atmospheric escape approach is widely used to estimate mass-loss rates for a broad range of planets that host hydrogen-dominated atmospheres as well as for performing atmospheric evolution calculations.}
   {We aim to study the applicability range  of the energy-limited atmospheric escape approximation.}
   {We revise the energy-limited atmospheric escape formalism and the involved assumptions. We also compare the results of the energy-limited formalism with those of hydrodynamic simulations, employing a grid covering planets with masses, radii, and equilibrium temperatures ranging between 1\,$M_{\oplus}$ and 39\,$M_{\oplus}$, 1\,$R_{\oplus}$ and 10\,$R_{\oplus}$, and 300 and 2000\,K, respectively.}
   {Within the grid boundaries, we find that the energy-limited approximation gives a correct order of magnitude estimate for mass-loss rates for about 76\% of the planets, but there can be departures from hydrodynamic simulations by up to two to three orders of magnitude in individual cases. Furthermore,  we find that planets for which the mass-loss rates are correctly estimated by the energy-limited approximation to within one order of magnitude have intermediate gravitational potentials ($\approx$2.5--5.5 $\times$10$^8$\,J\,kg$^{-1}$) as well as low-to-intermediate equilibrium temperatures and irradiation fluxes of extreme ultraviolet and X-ray radiation. However, for planets with low or high gravitational potentials, or high equilibrium temperatures and irradiation fluxes, the approximation fails in most cases.}
   {The energy-limited approximation should not be used for planetary evolution calculations that require computing mass-loss rates for planets that cover a broad parameter space. In this case, it is very likely that the energy-limited approximation would at times return mass-loss rates of up to several orders of magnitude above or below those predicted by hydrodynamic simulations. For planetary atmospheric evolution calculations, interpolation routines or approximations based on grids of hydrodynamic models should be used instead.}


   \maketitle
%

\section{Introduction}

It is believed that planets accrete an extensive hydrogen atmosphere when they form inside the protoplanetary nebula \citep[e.g.][]{stokl2016dynamical,lammer2018origin,owen2020}. Following the dispersion of the nebula, the hydrogen atmosphere can be subject to significant mass loss. In the beginning, the escape is driven by a combination of high temperatures and low gravity, leading to a phase of exhaustive hydrodynamic escape \citep[boil-off or core-powered mass loss;][]{stoekl2015hydrodynamic,owen2016atmospheres, ginzburg2016super,fossati2017aeronomical,kubyshkina2018grid, gupta2019,gupta2020}. The high temperatures in this phase of evolution are mostly due to energy released from the contracting planetary interior and, for the most close-in planets, to the high stellar UV, optical, and infrared irradiation. The low gravity, however, is due to the large planetary radius that, at the time of the dispersal of the protoplanetary nebula, reaches as far as the Hill radius \citep[e.g.][]{stoekl2015hydrodynamic,owen2016atmospheres,ginzburg2016super}.

Following this initial phase of extreme mass loss, atmospheric escape is no longer dominated by the thermal energy and low gravity of the planet, but rather by upper atmospheric heating caused by absorption of the stellar high-energy radiation, that is, extreme X-ray and ultraviolet (XUV) radiation \citep{hunten1982stability, lammer2003atmospheric,yelle2004aeronomy}. As the XUV luminosity of the host star declines with time following the evolution of (i.e. decrease in) the stellar rotation rate \citep{ribas2005evolution,mamajek2008improved,wright2011stellar,tu2015,johnstone2020}, the XUV-driven hydrodynamic escape weakens until other escape processes, such as Jeans escape or non-thermal escape processes, start to dominate mass loss.

The discovery of a large variety of planets, particularly those with radii in between those of Earth and Neptune \citep[e.g.][]{mullally2015planetary}, and the detection of structures in the observed exoplanet population, such as the radius gap separating super-Earths and mini-Neptunes at roughly 1.6 Earth radii \citep{fulton2017california}, demonstrate the importance of studying and understanding planetary atmospheric evolution. Such studies gain importance considering that the Solar System planets, including the Earth, have most likely gone through substantial atmospheric evolution as well \citep[e.g.][]{lammer2018origin,lammer2020}. Within this context, it has been shown that planetary atmospheric escape is not only a fundamental process shaping the observed exoplanet population, but also plays a key role in determining whether a planet develops a habitable environment \citep[e.g.][]{jin2014,owen2017,jin2018,owen2018,lammer2018origin,ginzburg2018,gupta2019,gupta2020}.

Different hydrodynamic models, such as those of \citet{yelle2004aeronomy}, \citet{garcia2007}, \citet{murray2009atmospheric}, \citet{koskinen2013}, \citet{salz2015high}, \citet{erkaev2016euv}, \citet{guo2016influence}, and \citet{kubyshkina2018grid}, have been used to study upper atmospheres, thermal mass loss, and how escape varies as a function of system parameters for planets hosting hydrogen-dominated atmospheres. However, these hydrodynamic models are rather complex, which makes them not particularly suitable for atmospheric evolution simulations. To overcome this, analytic approximations, in particular the energy-limited escape approximation \citep{Watson1981,erkaev2007roche}, are often used. 

Recently, \citet{kubyshkina2018grid} and \citet{kubyshkina2018overcoming} presented analytic functions for estimating mass-loss rates based on the outcome of hydrodynamic simulations that had already been successfully used for planetary evolution calculations \citep[e.g.][]{kubyshkina2019a,kubyshkina2019b,kubyshkina2020,modirrousta2020a,modirrousta2020b}. Nevertheless, the energy-limited approximation is still widely used by the community for estimating mass-loss rates for planets over a very broad parameter space \citep[e.g.][]{locci2019,poppenhaeger2021}. Therefore, we investigate here in detail the applicability of the energy-limited escape approach across a wide parameter space. 

Other authors have previously studied this applicability by means of comparisons with hydrodynamic simulations. \citet{kubyshkina2018overcoming} compared the results obtained from a large sample of hydrodynamic models (the same used in this study) with the energy-limited mass-loss rates. They reported that the energy-limited approximation significantly underestimates mass-loss rates for planets that are very close to boil-off conditions and moderately overestimates mass-loss rates for planets with (almost) hydrostatic atmospheres. However, in contrast to our present study, \citet{kubyshkina2018overcoming} employed an XUV absorption radius inferred from the simulations instead of that derived from the energy-limited formalism. \citet{murray2009atmospheric} followed an approach similar to that of \citet{kubyshkina2018overcoming} but focused on hot Jupiters and employed a less extensive set of hydrodynamic models. They observed that the energy-limited approximation leads to a significant overestimation of the mass-loss rates, by up to two orders of magnitude for highly irradiated hot Jupiters. They concluded that this difference is due to the fact that the energy-limited approximation wrongly estimates the level of unit optical depth to XUV radiation and that conductive cooling is irrelevant in the upper atmospheres of hot gas giants. \citet{erkaev2013xuv} found that the energy-limited mass-loss rates significantly overestimate the escape, particularly for planets with low irradiation levels, as they fail to meet the conditions of a hydrodynamically expanding atmosphere. Finally, \citet{tian2005transonic} suggested that the discrepancies between the energy-limited and the hydrodynamically computed mass-loss rates lie in the single-layer absorption assumption of the energy-limited approximation, which in many cases is not sufficiently accurate.
 
We aim to provide an assessment of the applicability of the energy-limited approximation over a wide parameter space, highlighting the regions characterised by large discrepancies between energy-limited and hydrodynamically computed mass-loss rates.\ We additionally attempt to provide reasons for the failure of the energy-limited approximation. Furthermore, in contrast to all previous studies, we only use XUV absorption radii predicted within the energy-limited framework.

This paper is organised as follows. We review the energy-limited formalism and the involved assumptions in Sect.~\ref{sec_ref}. In Sect.~\ref{sec_analysis} we present a comparison of the results obtained employing the simplest form of the energy-limited approximation with those obtained by solving the energy-limited equations. Section~\ref{sec_comp} presents a comparison of the energy-limited mass-loss rates with those obtained from hydrodynamic simulations. In Sect.~\ref{sec_disc} we review the applicability of the energy-limited approximation across the considered parameter space and draw our conclusions in Sect.~\ref{sec_concl}.

%
\section{Review of the energy-limited atmospheric escape approach}\label{sec_ref}
\citet{Watson1981} introduced the concept of energy-limited atmospheric escape in the context of hydrogen atmospheres. The aim was to provide an upper limit for the mass-loss rate by equating the escape flux and the incident XUV energy. \citet{erkaev2007roche} then extended the formulation of \citet{Watson1981} by considering tidal effects, obtaining
\begin{equation}
    \dot{M} = \pi \ \frac{\nu \ \Phi_{\rm XUV} \ r^2_{\rm XUV} \ r_0} {K \ G \ M_{\rm pl}}\,,
    \label{eq_escape}
\end{equation}
where $\dot{M}$ is the mass-loss rate, $\nu$ the efficiency of XUV heating \citep[see e.g.][]{shematovich2014heating,Salz2016}, $\Phi_{\rm XUV}$ the incident stellar XUV flux, $r_{\rm XUV}$ the radius at which the XUV flux is absorbed in the atmosphere, $r_{\rm 0}$ an arbitrary lower boundary radius usually defined as the photospheric radius of the planet ($R_{\rm pl}$), $G$ the gravitational constant, and $M_{\rm pl}$ the planetary mass. Here, $K$ is the correction factor, ranging between zero and one, which was introduced by \citet{erkaev2007roche} to account for the tidal effects that were first discussed by \citet{gu2003effect}. 
The existence of the Roche lobe means that to escape the gravitational potential of a planetary body, atmospheric particles must only reach the Roche lobe, which in turn means that the energy each individual particle needs to escape is lower than if no host star were present. The correction factor $K$, which is particularly important for close-in low-mass planets, is defined as \citep{erkaev2007roche}
\begin{equation}
    K = 1 - \frac{3 R_{\rm pl}}{2R_{\rm Rl}} + \frac{R_{\rm pl}^3}{2 R_{\rm Rl}^3}\,,
\end{equation}
where the Roche radius ($R_{\rm Rl}$) is
\begin{equation}\label{eq.roche_radius}
    R_{\rm Rl} \approx d_0 \left( \frac{M_{\rm pl}}{3 M_{\rm star}} \right)^{\frac{1}{3}}\,.
\end{equation}
In Eq.~(\ref{eq.roche_radius}), $d_0$ is the planetary orbital separation and $M_{ \rm star}$ is the stellar mass.

The term $r_{\rm XUV}$ is usually unknown and, particularly in extended hydrogen envelopes, is significantly larger than $r_0$ \citep{Watson1981,kubyshkina2018grid}. Employing a variety of assumptions and simplifications, detailed below, \citet{Watson1981} derived a system of equations that allows the simultaneous calculation of $\dot{M}$ and $r_{\rm XUV}$ with numerical methods. Below we summarise the framework of \citet{Watson1981}.

We remark that throughout this work we consider the atmospheric mass-loss rate ($\dot{M}$) in kg\,s$^{-1}$, while \citet{Watson1981} primarily considered the particle escape rate ($F$) in particles\,sr$^{-1}$\,s$^{-1}$. The relation between $F$ and $\dot{M}$ is
\begin{equation}
    \dot{M} = 4 \ \pi \ m \ F\,.
\end{equation}
\subsection{Model description and assumptions}\label{sec:assumptions}
\citet{Watson1981} defined dimensionless variables for atmospheric height ($\lambda$), temperature ($\tau$), velocity ($\Psi$), particle escape ($\zeta$), and the energy parameter ($\beta$), which, accounting for tidal effects, are defined as
\begin{equation}
\label{eq_lambda}
    \lambda = \frac{K \ G \ M_{\rm pl} \ m}{k \ T_0 \ r}\,,
\end{equation}
\begin{equation}
\label{eq_tau}
    \tau = \frac{T}{T_0}\,,
\end{equation}
\begin{equation}
\label{eq_psi}
    \Psi = \frac{m \ u^2}{k \ T_0}\,,
\end{equation}
\begin{equation}
\label{eq_zeta}
    \zeta = \dot{M} \ \frac{ \ k^2 \ T_0}{4\pi \kappa_0 \ K \ G \ M_{\rm pl} \ m^2}\,,
\end{equation}
and
\begin{equation}
\label{eq_beta}
    \beta = \nu \ \Phi_{\rm XUV} \ \frac{K \ G \ M_{\rm pl} \ m}{k \ T_0^2 \ \kappa_0}\,,
\end{equation}
respectively. In Eqs.~(\ref{eq_lambda}) to (\ref{eq_beta}), $k$ is the Boltzmann constant, $m$ the mass of an escaping particle, $T_0$ the temperature at $r_0$, $\kappa_0$ the thermal conductivity at $r_0$, $r$ an arbitrary radius at or above $r_0$, $T$ the temperature at $r$, and $u$ the bulk velocity at $r$. We note that $\lambda$ is the Jeans escape parameter and that, when using the photospheric radius $R_{ \rm pl}$ as $r$ and the equilibrium temperature $T_{ \rm eq}$ as $T_0$, Eq.~(\ref{eq_lambda}) reduces to the `restricted Jeans escape parameter' \citep{fossati2017aeronomical}:
\begin{equation}
    \Lambda = \frac{K \ G\ M_{ \rm pl} \ m}{k \ T_{ \rm eq} \ R_{ \rm pl}}\,.
\end{equation}

\citet{Watson1981} parametrised the thermal conductivity of a neutral gas as
\begin{equation}
    \kappa = \kappa_0 \ \tau^{s}\,,
\end{equation}
where $s = 0.7$ is usually chosen for a hydrogen-dominated gas. Using Eqs.~(\ref{eq_lambda}), (\ref{eq_tau}), and (\ref{eq_psi}), the steady state equations for mass, momentum, and energy conservation \citep[see Eqs. 3, 4, and 5 in][]{Watson1981} can be reduced to the following two equations
\begin{equation}
    \left( 1 - \frac{\tau}{\Psi} \right) \frac{d \Psi}{d \lambda} = 2 \left( 1- \frac{2\tau}{\lambda} - \frac{d\tau}{d\lambda} \right)
    \label{eq_Watson1}
\end{equation}
and
\begin{equation}
    \frac{\tau^{s}}{\zeta} \frac{d\tau}{d\lambda} = \epsilon + \lambda - \frac{5 \tau}{2} - \frac{\Psi}{2}\,,
    \label{eq_Watson2}
\end{equation}
where $\epsilon$ is the energy each escaping particle carries away from the system and is defined as
\begin{equation}\label{eq.epsilon}
    \epsilon = \epsilon_{\infty} - \frac{k}{\zeta \ \kappa_0 \ K \ G \ M_{\rm pl} \ m} \int_r^{\infty} q \ r^2 \ dr\,.
\end{equation}
In Eq.~(\ref{eq.epsilon}), $q$ is the volume heating rate and $\epsilon_{\rm \infty}$ denotes the energy flow at infinity. In the special case where $\epsilon_{\rm \infty} = 0$, the energy flowing outwards is just enough to lift the gas from the gravitational field but does not leave any excess energy at very large distances. \citet{Watson1981} assumed this to be the case for every escaping particle since any additional energy carried away would only decrease the maximum particle escape flux. Therefore, $\epsilon_{\rm \infty}$ is set equal to zero. 

Three assumptions need to be made to define $\epsilon$: 1) The incident XUV energy is absorbed in a single narrow atmospheric layer; 2) the fraction of the absorbed radiation spent on escape is equal to the heating efficiency ($\nu$); and 3) the incident XUV energy is averaged over a sphere by a factor of 1/4 as this is the ratio of the cross-section and surface area of a sphere \citep{erkaev2007roche}. With these assumptions, Eq.~(\ref{eq.epsilon}) becomes
\begin{equation}
    \epsilon =  - \frac{\pi \ m \ \nu \ \Phi_{\rm XUV} \ r_{\rm XUV}^2}{\dot{M} \ k \ T_0}\,.
\end{equation}
Employing Eqs.~(\ref{eq_zeta}) and (\ref{eq_beta}), the hydrodynamic equations (\ref{eq_Watson1}) and (\ref{eq_Watson2}), valid for the atmospheric region lying below $r_{\rm XUV}$, can be further reduced to \citep{Watson1981}
\begin{equation}
\label{eq_Watson3}
    \frac{\tau^{s}}{\zeta}\frac{d\tau}{d\lambda} = \lambda - \frac{\beta}{\zeta \lambda_{\rm XUV}^2} - \frac{5\tau}{2} - \frac{\Psi}{2}\,.
\end{equation}
By assuming $r_{\rm XUV}$ to be well below the sonic level, namely where the velocity of the gas reaches the sound speed, the velocity term can be set equal to zero. Furthermore, close to the lower boundary, $\lambda >> \tau$. With these additional considerations, Eq.~(\ref{eq_Watson3}) can be further simplified to \citep{Watson1981}
\begin{equation}
\label{eq_watson4}
    \frac{\tau^{s}}{\zeta}\frac{d\tau}{d\lambda} = \lambda - \frac{\beta}{\zeta \lambda_{\rm XUV}^2}\,.
\end{equation}

Finally, \citet{Watson1981} made assumptions regarding the shape of the temperature profile in the atmosphere. They argued that it has to be such that the escaping energy, the downward conduction of energy, and adiabatic cooling are in balance. A maximum particle escape rate can therefore be obtained by allowing the temperature in the thermosphere to drop to a 0\,K minimum somewhere between the lower boundary and the XUV absorption region. This is because in this case the least possible amount of the absorbed energy is lost to downward conduction and the largest possible amount of absorbed energy is used to fuel the escape. Then, the fact that at the 0\,K minimum both $\tau$ and $\frac{d \tau}{d \lambda}$ must be zero leads to the derivation of a system of equations that allows one to simultaneously calculate the escape parameter ($\zeta$) and the Jeans escape parameter at the absorption height ($\lambda_{\rm XUV}$) for this special temperature profile as \citep{Watson1981}
\begin{equation}
\label{eq_energylimited1}
    \zeta = \frac{2}{s+1} \left[ \frac{ \left( \frac{\lambda_{\rm XUV}}{2} \right)^{\frac{s+1}{2}} + 1 }{\Lambda - \lambda_{\rm XUV}} \right]^2
\end{equation}
and
\begin{equation}
\label{eq_energylimited2}
    \lambda_{\rm XUV} = \sqrt{\frac{\beta}{\zeta} \left[ \Lambda - \sqrt{ \frac{2}{(s+1) \zeta}} \right]^{-1}}\,.
\end{equation}

Below we summarise the major assumptions and simplifications employed by \citet{Watson1981} to derive Eqs.~(\ref{eq_energylimited1}) and (\ref{eq_energylimited2}).

\subsubsection{Hydrogen-dominated atmosphere} The atmosphere is assumed to be dominated by hydrogen, meaning that above $r_0$ the mixing ratios of species heavier than hydrogen must be significantly smaller than one. \citet{Watson1981} used an arbitrarily defined lower boundary height ($r_0$) that lies well above the homopause, but in an atmospheric region that is still dense enough to have an optical depth to XUV much larger than one. However, as long as hydrogen is the dominant atmospheric gas, the extent of the heterosphere is negligible in comparison to the extent of the whole atmosphere, allowing one to use $R_{\rm pl}$ as the lower boundary height ($r_0$). In planetary systems with extensive heterospheres, the energy-limited equations may only be used in regions well above the homopause.

\subsubsection{Hydrodynamic escape} The XUV-driven hydrodynamic escape is assumed to be the dominating escape mechanism in the atmosphere. Other escape processes, such as Jeans escape or non-thermal escape processes (e.g. sputtering, ion pickup, and charge exchange), must be negligible. The hydrodynamic escape must also be fuelled entirely by the incident XUV energy of the host star.

\subsubsection{Narrow absorption region} All the incident XUV energy is assumed to be absorbed in a narrow region near $r_{\rm XUV}$. Furthermore, the optical depth to XUV radiation at $r_{\rm XUV}$ is equal to one. 

\subsubsection{Gas properties} The gas is assumed to be non-viscous and to have a constant mean molecular weight. The gas pressure is assumed to be isotropic and to decline towards zero at infinity. \citet{Watson1981} argued that allowing for the pressure at infinity to remain above zero would decrease the maximum particle escape rate because it applies an additional force working against the outward expanding gas.

\subsubsection{No excess energy} The escaping particles are assumed to carry no excess energy away from the planet. This implies that each individual particle gets just enough energy to reach escape velocity, but no more. \citet{Watson1981} argued that additional excess energy would decrease the maximum particle escape rate because it would decrease the efficiency of XUV-driven hydrodynamic escape.

\subsubsection{Tightly bound condition} The atmosphere at and below the lower boundary ($r_0$) is assumed to be `tightly bound'. This terminology, which was first introduced by \citet{parker1964dynamical}, means that the atmosphere is quasi-static and expands subsonically upwards. Conversely, if the atmosphere is not tightly bound, the thermal velocity of the gas is comparable to or higher than the escape velocity, leading to supersonic catastrophic escape. \citet{Watson1981} defined the tightly bound condition in accordance with \citet{parker1964dynamical} to be met in all systems with $\Lambda \gtrsim 10$. \citet{volkov2011thermally} showed that catastrophic escape sets in at values of $\Lambda < 2.4$, but that significant deviations from tightly bound models can already be observed at $\Lambda < 6$.

\subsubsection{Subsonic assumption}The XUV absorption height ($r_{\rm XUV}$) is assumed to be below the sonic level. This assumption enables one to neglect the velocity term in Eqs.~(\ref{eq_Watson1}) and (\ref{eq_Watson2}) and to further place limits to the temperature in the thermosphere. \citet{Watson1981} argued that this assumption may break down for strongly heated planets with high mass-loss rates.
\subsection{Thermospheric temperature profile}
\label{sec_temperature}
The atmospheric temperature profile obtained through the energy-limited escape approach is one of the keys needed to understand the framework's validity and limitations. As mentioned above, the energy-limited approach assumes that the temperature in the thermosphere drops to a 0\,K minimum before the atmosphere is reheated by the absorbed incident XUV radiation. The energy acquired by XUV absorption is then redistributed within the gas as that heat is then used to drive the escape. We summarise here the mathematical description of the temperature profile in the region between the lower boundary and the XUV absorption height ($r_{\rm XUV}$).

The starting point is Eq.~(\ref{eq_watson4}), which is a differential equation that can be solved via integration. The conditions $\lambda = \Lambda$ and $\tau = 1$ are used for the lower boundary, while for the upper boundary the arbitrary values $\lambda'$ and $\tau'$ are used. The solution is obtained through the following steps
\begin{equation}
    \tau^{s} d\tau = \bigg(\lambda \zeta - \frac{\beta}{\lambda^2_{\rm XUV}}\bigg) d\lambda\,,
\end{equation}
\begin{equation}
\label{eq:integration}
    \int_{ \rm 1}^{\tau'} \tau^{s} d\tau = \int_{ \rm \Lambda}^{\lambda'} \bigg(\lambda \zeta - \frac{\beta}{\lambda^2_{\rm XUV}}\bigg) d\lambda\,,
\end{equation}
\begin{equation}
    \frac{\tau^{s+1}}{s+1} \bigg\vert_{ 1}^{\tau'} = \bigg(\frac{\lambda^2}{2} \zeta - \frac{\beta}{\lambda_{\rm XUV}^2} \lambda \bigg) \bigg\vert_{ \rm \Lambda}^{\lambda'}\,,
\end{equation}
\begin{equation}
    \frac{\tau'^{s+1}}{s+1} - \frac{1}{s+1} = \bigg( \frac{\lambda'^2}{2} \zeta - \frac{\beta}{\lambda_{\rm XUV}^2} \lambda' \bigg) - \bigg( \frac{\Lambda^2}{2} \zeta - \frac{\beta}{\lambda_{\rm XUV}^2} \Lambda \bigg)\,,
\end{equation}
and
\begin{equation}\label{eq.lastT}
    \frac{\tau'^{s+1} - 1}{s+1} = \frac{\zeta}{2} \bigg( \lambda'^2 - \Lambda^2 \bigg) - \frac{\beta}{\lambda_{ \rm XUV}^2} \bigg( \lambda' - \Lambda \bigg)\,.
\end{equation}
Solving Eq.~(\ref{eq.lastT}) for $\tau'$ yields the atmospheric temperature as a function of height in the atmosphere ($\lambda'$). For the sake of simplicity, all primed values can be replaced with unprimed values, resulting in%
\begin{equation}
    \tau = \bigg\{ 1 +  (s+1) \bigg[ \frac{\zeta}{2} (\lambda^2 - \Lambda^2 ) - \frac{\beta}{\lambda_{ \rm XUV}^2} (\lambda - \Lambda) \bigg] \bigg\}^{\frac{1}{s+1}}\,.
    \label{eq_temperatureprofile}
\end{equation}

The temperature profile we derived above is slightly different than that given by \citet{Watson1981}. The difference lies in the choice of the lower boundary used for the integration of Eq.~(\ref{eq:integration}) that we consider to be the lower atmospheric boundary (i.e. the planetary radius), while \citet{Watson1981} considered it to be the location of the 0\,K minimum. We modified the computation of the temperature profile for the following reasons: 1) The temperature profile described by Eq.~(\ref{eq_temperatureprofile}) allows for the calculation of the atmospheric temperature at the XUV absorption radius ($T_{\rm XUV}$), whether or not there is a 0\,K minimum within the thermosphere; 2) the $T_{\rm XUV}$ values computed by employing Eq.~(\ref{eq_temperatureprofile}) are physically more reasonable compared to those computed with the equation derived by \citet{Watson1981}, which in several cases predicts  $T_{\rm XUV}$ values reaching as high as 10$^5$\,K; 3) finally, Eq.~(\ref{eq_temperatureprofile}) takes the physical conditions of the atmosphere at the lower boundary into account, in contrast to the equation derived by \citet{Watson1981}. We remark that Eq.~(\ref{eq_temperatureprofile}) meets all the assumptions of the energy-limited formalism.

If the height of the XUV absorption ($\lambda_{\rm XUV}$) is known or fixed, Eq.~(\ref{eq_temperatureprofile}) can be used to investigate the dependence of the escape rate on the temperature profile. As long as the escape flux is below a critical value, the temperature will monotonically rise between $\Lambda$ and $\lambda_{\rm XUV}$. If the escape flux increases further, the remaining XUV energy, which is conducted down to the lower atmosphere, is not sufficient to counteract adiabatic cooling, and a temperature minimum will appear \citep{Watson1981}.

Equation~(\ref{eq_temperatureprofile}) analytically describes the temperature profile only if $\lambda_{\rm XUV}$ is known, which therefore must be given as input. We remark that Eq.~(\ref{eq_temperatureprofile}) is strongly dependent on the assumption that the height of the XUV absorption lies well below the sonic level, and thus the derived temperature profile is valid only below that point.
\section{Consequences of approximating the location of the XUV absorption radius}\label{sec_analysis}
Having described the details and assumptions involved in the energy-limited escape approach, we analyse now the resulting planetary properties (escape rate, absorption height, pressure, and temperature) obtained from applying it as a function of planetary mass, radius, and incident stellar XUV flux. To this end, we employed about 250,000 synthetic planets extracted from a planetary grid, whose characteristics are listed in Table~\ref{tb_grid}. From this grid, we then considered only planets complying with the following four conditions: 1) $10 \leq \Lambda \leq 200$, 2) average density between 30 and 10$^4$\,kg\,m$^{-3}$, 3) $R_{\rm Rl}$ $>$ 1.5\,$R_{\rm pl}$, and 4) $\log(-\Phi_G) = \log\left(\frac{G M_{ \rm pl}}{R_{ \rm pl}} \right) \leq 9.11$, in accordance with \citet{Salz2016}, who found that planets with higher gravitational potential become hydrodynamically stable.

For all calculations, we adopted a heating efficiency of $\nu = 0.15$ \citep{shematovich2014heating,Salz2016,kubyshkina2018grid}. We remark that the question of what $\nu$ value to adopt is still strongly debated. Attempts to constrain it for a range of planets have been performed for example by \citet{shematovich2014heating} and \citet{Salz2016}. These studies have found that, although slight variations are to be expected, $\nu$ remains around a value of roughly 20\% for the planets considered in this work. However, the aim of this paper is not to provide an algorithm for calculating mass-loss rates, which would require an accurate estimate of $\nu$ for each individual planet, but rather to compare the results of the energy-limited approximation with results of hydrodynamic simulations, both of which require $\nu$ as input. Therefore, we chose a $\nu$ to match the one employed to compute the hydrodynamic simulations.
\begin{table*}[ht!]
\centering
\caption{Properties of the grid employed to study the behaviour of the energy-limited approach described in Sect.~\ref{sec_analysis}.}
\label{tb_grid}
\begin{tabular}{c|c|c|c|c}
\hline\hline   
     Parameter &  Lower Limit & Upper Limit & Number of grid nodes & Spacing\\ \hline
     $M_{\rm pl}$ & $0.5 M_{\rm \oplus}$ & $318 M_{\rm \oplus} \approx 1 M_{\rm \jupiter}$ & 80 & geometric \\ \hline
     $R_{\rm pl}$ & $1 R_{\rm \oplus}$ & $11.2 R_{\rm \oplus} \approx 1 R_{\rm \jupiter}$ & 28 & linear \\ \hline
     $T_{\rm eq}$ & 300\,K & 2000\,K & 18 & linear \\ \hline
     $L_{\rm XUV}$ & 4.75$\times$10$^{20}$\,J\,s$^{-1}$ & 1.64$\times$10$^{23}$\,J\,s$^{-1}$ & 20 & geometric \\ \hline
\end{tabular}
\tablefoot{$M_{\rm pl}$ is the planetary mass, $R_{\rm pl}$ is the planetary radius, $T_{\rm eq}$ is the planetary equilibrium temperature, and $L_{\rm XUV}$ is the XUV luminosity of the host star. A geometric spacing indicates that the grid points have been set by employing a geometric progression.}
\end{table*}

\begin{figure}[ht!]
\centering
\includegraphics[width=\columnwidth]{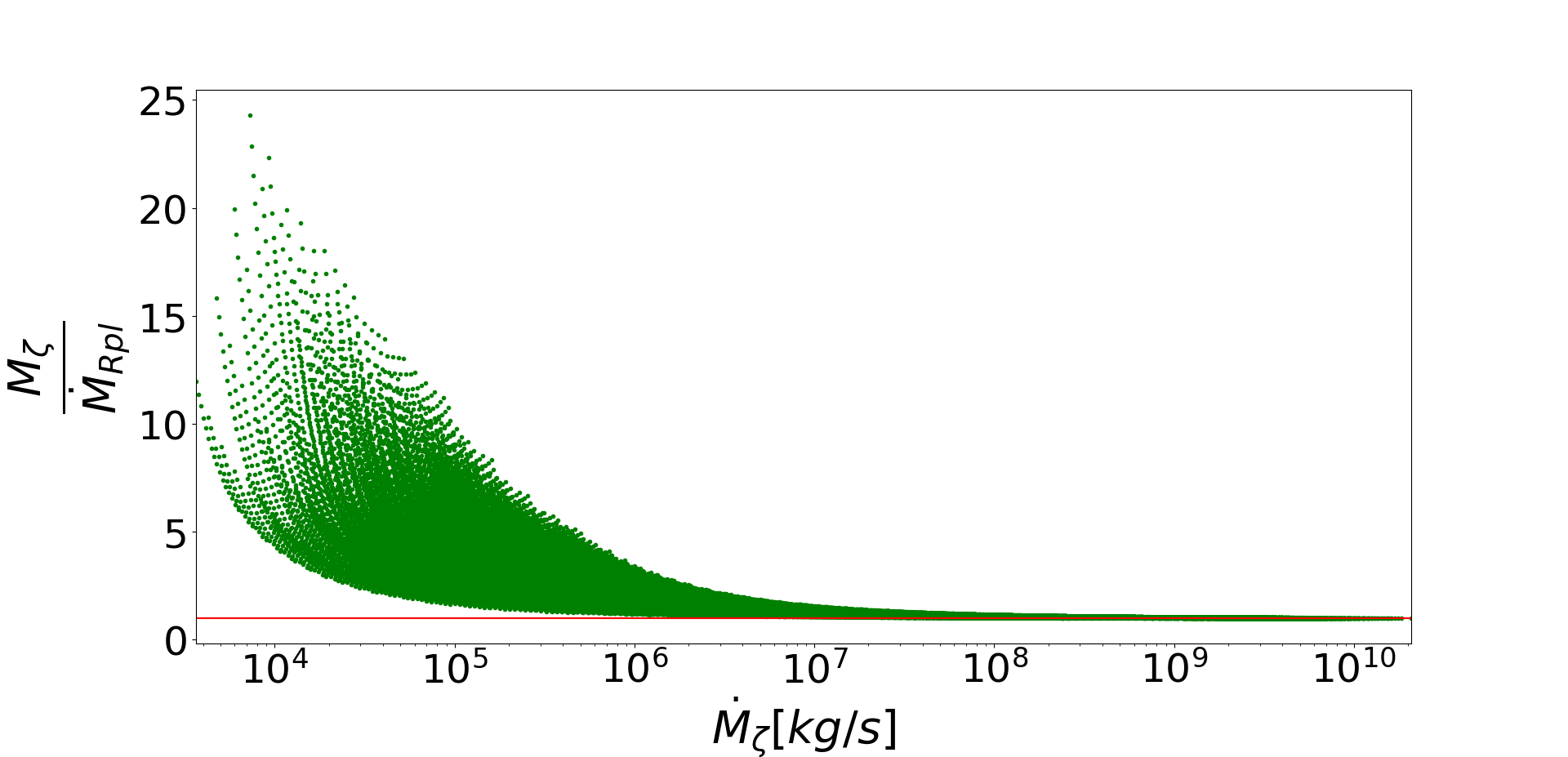}
\includegraphics[width=\columnwidth]{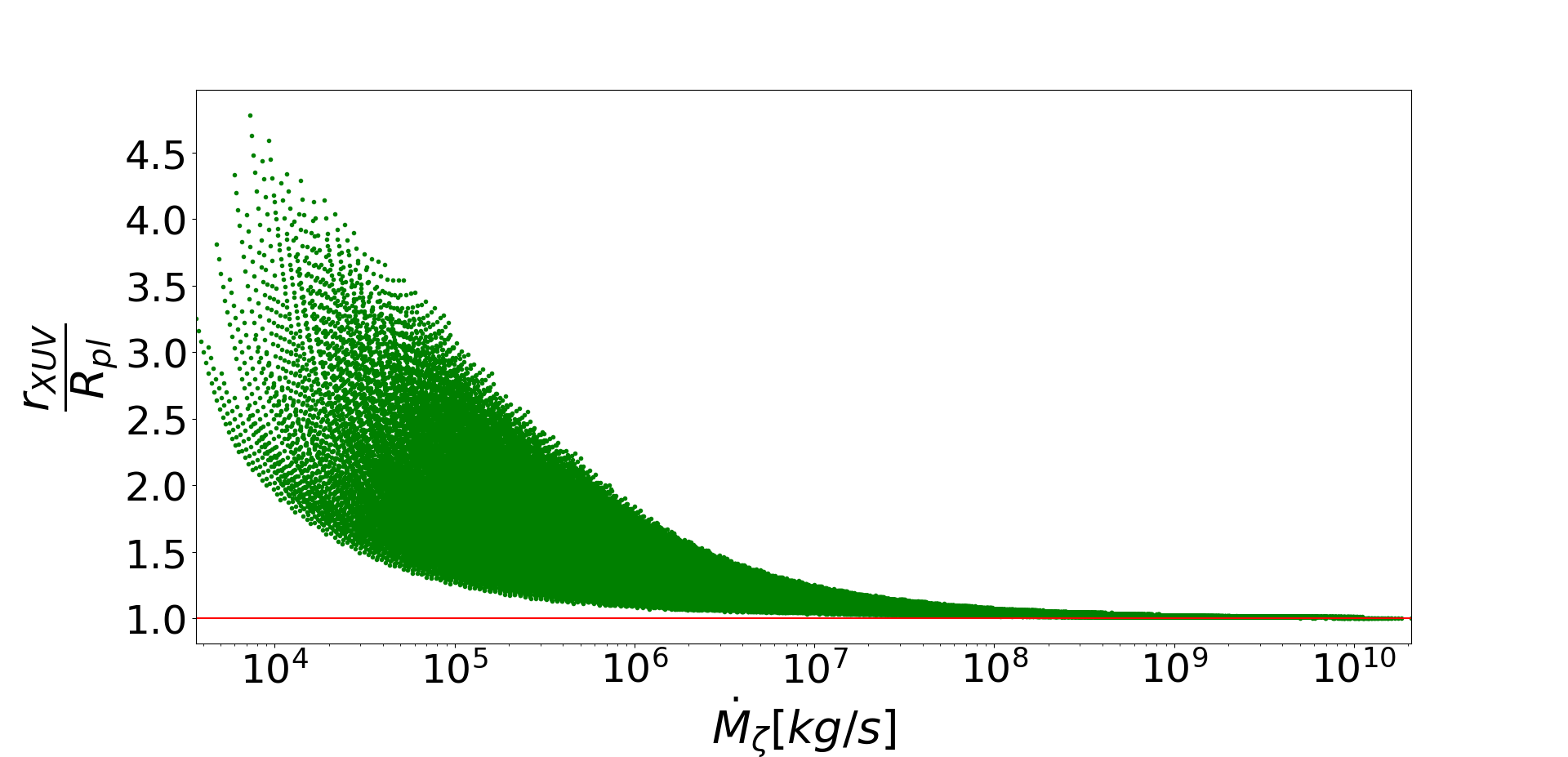}
\includegraphics[width=\columnwidth]{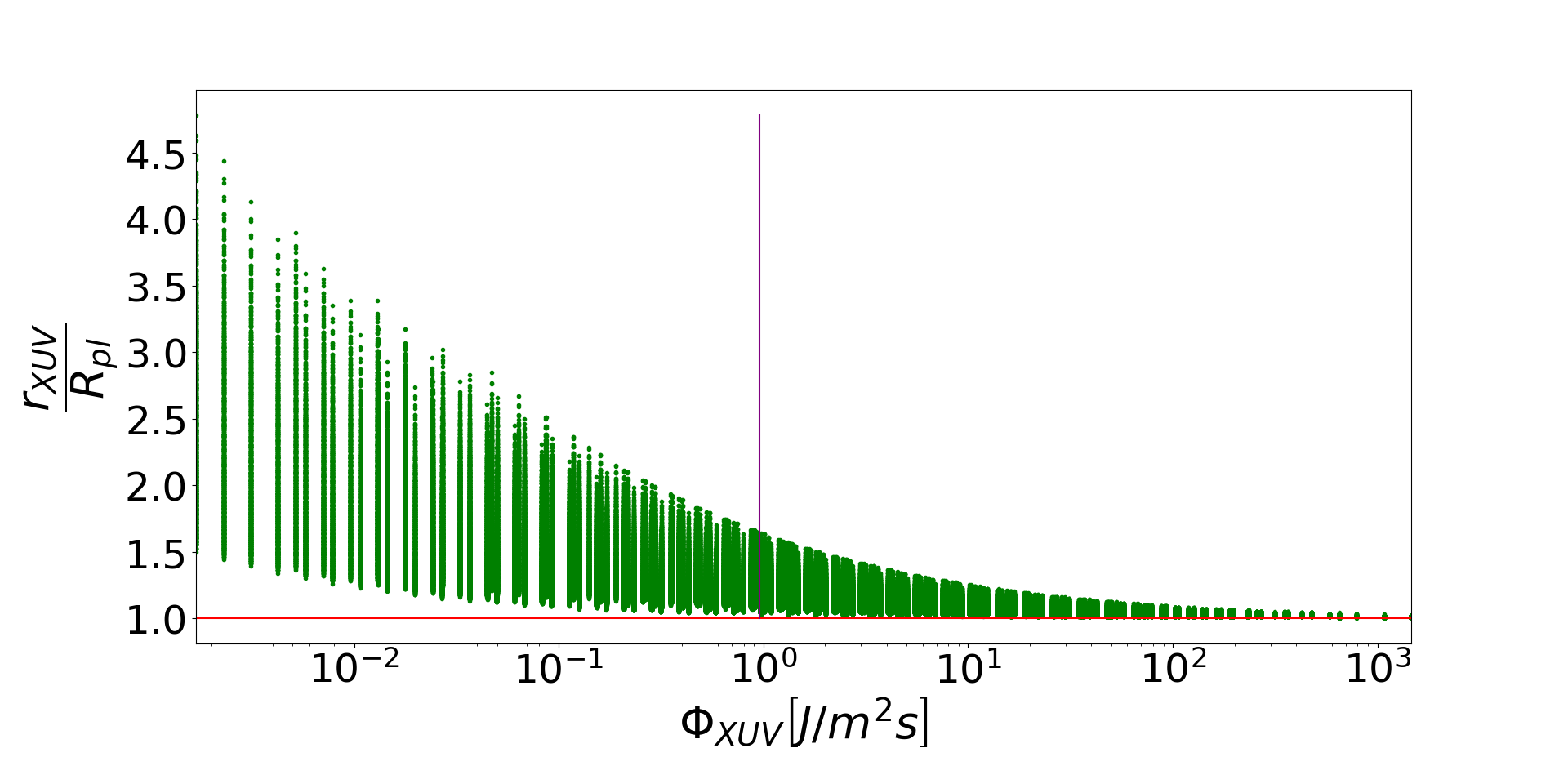}
\caption{Ratios of calculated mass-loss rates and XUV absorption radii using different versions of the energy-limited approximation. Top: Ratio between the mass-loss rate obtained by numerically solving Eqs.~(\ref{eq_energylimited1}) and (\ref{eq_energylimited2}) ($\dot{M}_{\rm \zeta}$) and the mass-loss rate obtained through Eq.~(\ref{eq_escape_simple}) ($\dot{M}_{\rm R_{\rm pl}}$) as a function of $\dot{M}_{\rm \zeta}$. Middle: Same as the top panel, but for the ratio between $r_{\rm XUV}$, obtained through solving Eqs.~(\ref{eq_energylimited1}) and (\ref{eq_energylimited2}), and $R_{\rm pl}$. Bottom: Same as the middle panel, but as a function of the incident stellar XUV flux $\Phi_{\rm XUV}$. For reference, the vertical line shows the XUV flux incident on HD97658\,b \citep{kubyshkina2018overcoming}. In each panel, the red line marks a ratio of one for reference.}
\label{fig_zeta_vs_rpl}
\end{figure}

Often, the energy-limited approach is used to approximate the XUV absorption radius ($r_{ \rm XUV}$) with the planetary radius \citep[$R_{\rm pl}$; e.g.][]{ehrenreich2011,owen2017,dorn2018,jin2018}. With this additional approximation, the resulting energy-limited mass-loss rate is
\begin{equation}
    \dot{M}_{\rm R_{\rm pl}} = \pi \ \frac{\nu \ \Phi_{\rm XUV} \ R_{\rm pl}^3} {K \ G \ M_{\rm pl}}\,.
    \label{eq_escape_simple}
\end{equation}
The top panel of Fig.~\ref{fig_zeta_vs_rpl} shows the ratio of the mass-loss rate obtained by numerically solving Eqs.~(\ref{eq_energylimited1}) and (\ref{eq_energylimited2}) (i.e. $\dot{M}_{\rm \zeta}$) to the mass-loss rate obtained from Eq.~(\ref{eq_escape_simple}) (i.e. $\dot{M}_{\rm R_{pl}}$) and thus assuming $r_{\rm XUV}$\,=\,$R_{\rm pl}$ as a function of $\dot{M}_{\rm \zeta}$. For large values of $\dot{M}_{\rm \zeta}$ the two approaches lead to comparable mass-loss rates. However, $\dot{M}_{\rm R_{\rm pl}}$ underestimates the energy-limited mass-loss rate for values of $\dot{M}_{\rm \zeta}$ smaller than 10$^7$\,kg\,s$^{-1}$ by up to more than an order of magnitude.
The middle panel of Fig.~\ref{fig_zeta_vs_rpl} shows how this behaviour appears in terms of the ratio of the XUV absorption radius ($r_{\rm XUV}$) obtained with the energy-limited approach to the planetary radius ($R_{\rm pl}$).  

The bottom panel of Fig.~\ref{fig_zeta_vs_rpl} illustrates how the $r_{\rm XUV}$/$R_{\rm pl}$ radius ratio behaves as a function of incident stellar XUV flux. The two radii get closer as the stellar XUV flux increases. This means that the increase in the mass-loss rate with increasing $\Phi_{\rm XUV}$ is slowed down by the decreasing $r_{ \rm XUV}$.

\begin{figure}
\centering
\includegraphics[width=\columnwidth]{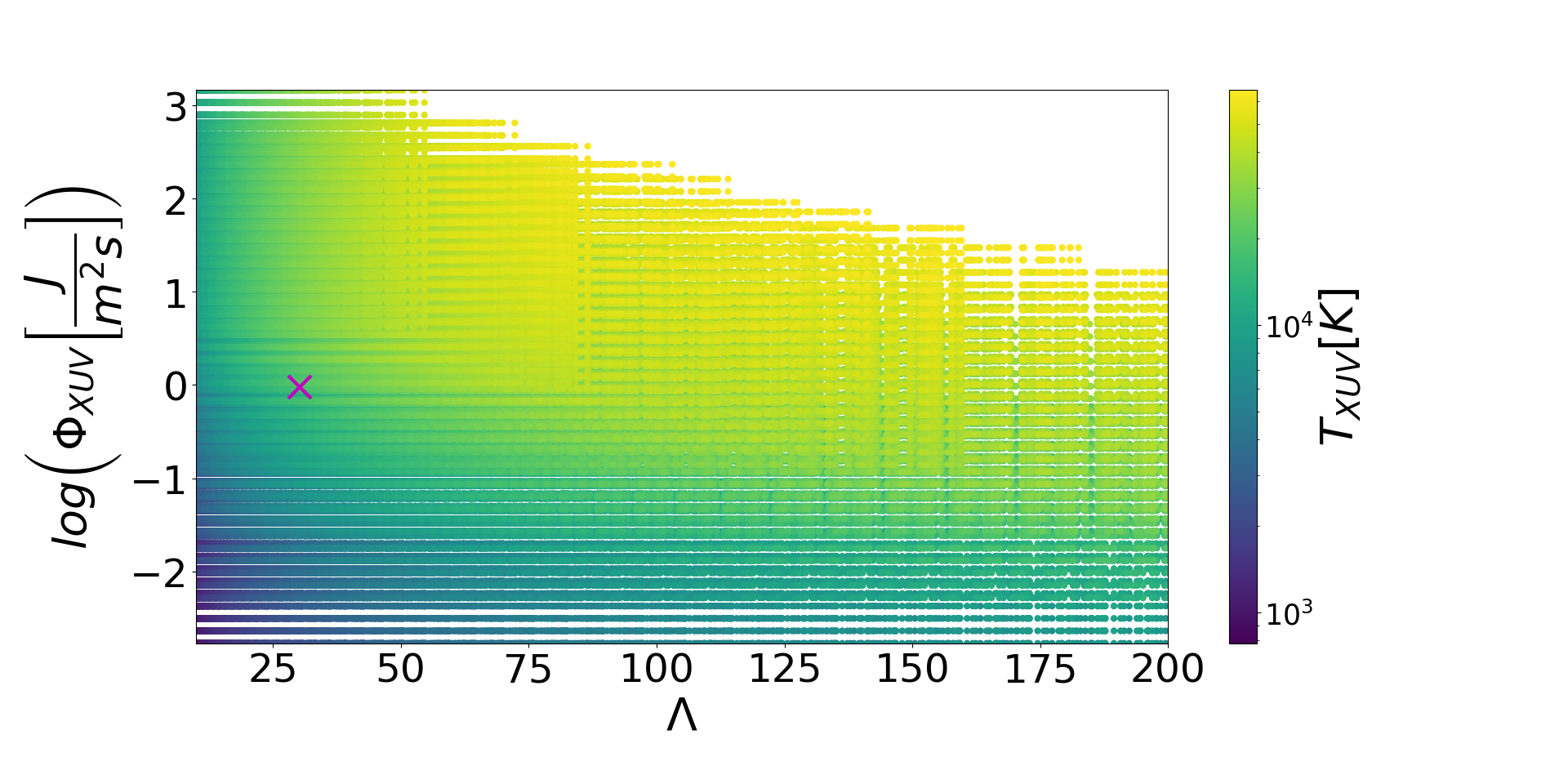}
\includegraphics[width=\columnwidth]{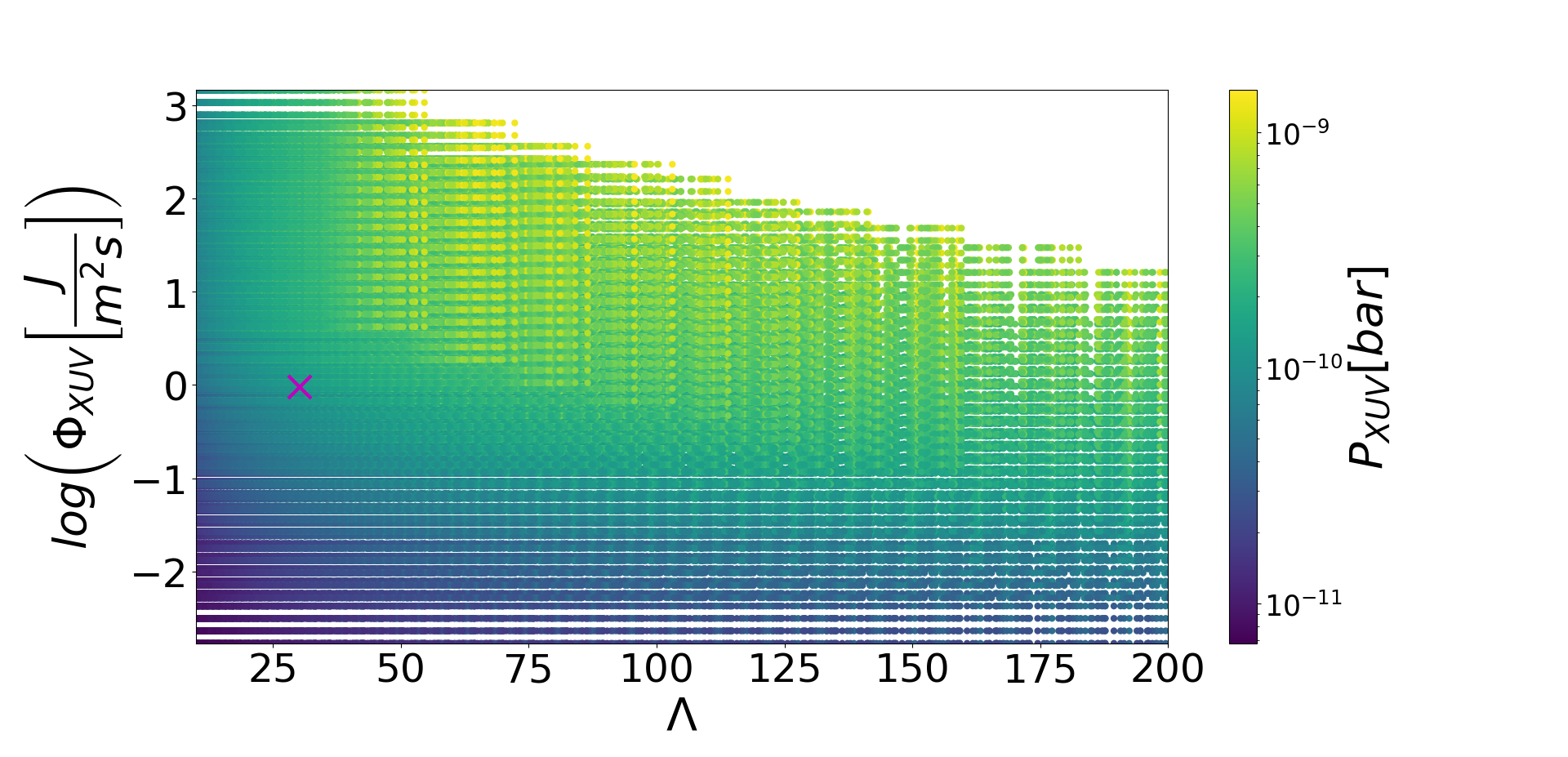}
\caption{Temperature (top) and pressure (bottom) at the XUV absorption layer as a function of $\Lambda$ and of the incident XUV flux. For reference, the purple X marks the position of HD97658\,b \citep{kubyshkina2018overcoming}.}
\label{fig_temperature}
\end{figure}

Figure~\ref{fig_temperature} shows the atmospheric temperature ($T_{\rm XUV}$) and pressure ($P_{\rm XUV}$) at the XUV absorption radius as a function of both stellar XUV flux and $\Lambda$. The $T_{\rm XUV}$ values range between 10$^3$ and 50,000\,K, and $T_{\rm XUV}$ increases with increasing stellar XUV flux and $\Lambda$. 

The $P_{\rm XUV}$ values are not directly obtained from the energy-limited formalism, but they can be derived as follows. In a static and isothermal atmosphere, the condition defining the level of unit optical depth to XUV radiation is \citep{Watson1981}
\begin{equation}
\label{eq_opticaldepth}
    n_{\rm XUV} \ H_{\rm XUV} \ \sigma_{\rm XUV} \approx 1\,,
\end{equation}
where
\begin{equation}
\label{eq_scaleheight}
    H_{\rm XUV} = \frac{k \ T_{\rm XUV} \ r_{\rm XUV}^2}{K \ G \ M \ m}
\end{equation}
is the scale height at the XUV absorption radius, $n_{\rm XUV}$ is the particle density at the XUV absorption radius, and $\sigma_{\rm XUV} \approx 2 \times 10^{-22}$\,m$^2$ is the absorption cross-section of atomic hydrogen to XUV radiation at an XUV photon energy of 20\,eV \citep{murray2009atmospheric}. We remark that, because the atmosphere is not static but accelerating upwards and because the temperature is declining from $r_{\rm XUV}$ upwards, the true scale height is smaller than that given by Eq.~(\ref{eq_scaleheight}), and thus Eq.~(\ref{eq_opticaldepth}) becomes \citep{Watson1981}
\begin{equation}
    n_{\rm XUV} \ H_{\rm XUV} \ \sigma_{\rm XUV} > 1\,.
\end{equation}
However, Eq.~(\ref{eq_opticaldepth}) can still be used to estimate the pressure at the XUV absorption height that Fig.~\ref{fig_temperature} shows lying in the 10$^{-11}$ to 10$^{-9}$\,bar range. The energy-limited approach indicates that $P_{\rm XUV}$ increases with increasing stellar XUV flux and $\Lambda$.
\section{Comparison with hydrodynamic simulations}\label{sec_comp}
We compare here the results obtained through the energy-limited approach with those derived from hydrodynamic atmospheric modelling. \citet{kubyshkina2018grid} presented the results of one-dimensional hydrodynamic simulations of hydrogen-dominated planetary atmospheres for about 7000 synthetic non-magnetised planets. By construction, the hydrodynamic simulations account for Jeans escape, XUV hydrodynamic escape, and boil-off (i.e. core-powered mass loss). They also developed an interpolation routine that allows one to derive mass-loss rates ($\dot{M}_{\rm hc}$) for planets lying within the grid boundaries.

We employed the energy-limited formalism to derive $\dot{M}_{\rm \zeta}$ for roughly 97,000 planets contained within a grid we constructed by employing boundaries equal to those of the grid built by \citet{kubyshkina2018grid}. The grid parameters are given in Table~\ref{tb_grid2}. Furthermore, this grid also complies with the conditions listed in Sect.~\ref{sec_analysis}, but we applied a stricter upper limit on $\Lambda$ (i.e. $\Lambda \leq 80.0$) to better match the grid of \citet{kubyshkina2018grid}. We remark that \citet{kubyshkina2018grid} performed a comparison similar to that shown here, but with the key difference that \citet{kubyshkina2018grid} computed $\dot{M}_{\rm \zeta}$ by employing $r_{\rm XUV}$ values extracted from the hydrodynamic simulations, while we employed the $r_{\rm XUV}$ values derived from the energy-limited formalism (i.e. solving Eqs.~(\ref{eq_energylimited1}) and (\ref{eq_energylimited2})).
\begin{table*}[t]
\centering
\caption{Same as Table~\ref{tb_grid}, but for the grid used in Sect.~\ref{sec_comp} to compare the results of the energy-limited approach with those of hydrodynamic simulations.}
\label{tb_grid2}
\begin{tabular}{c|c|c|c|c}
\hline\hline   
     Parameter &  Lower Limit & Upper Limit & Number of grid nodes & Spacing\\ \hline
     $M_{\rm pl}$ & $1 M_{\rm \oplus}$ & $39 M_{\rm \oplus}$ & 39 & linear \\ \hline
     $R_{\rm pl}$ & $1 R_{\rm \oplus}$ & $10 R_{\rm \oplus}$ & 10 & linear \\ \hline
     $T_{\rm eq}$ & 300\,K & 2000\,K & 18 & linear \\ \hline
     $L_{\rm XUV}$ & 4.75$\times$10$^{20}$\,J\,s$^{-1}$ & 1.64$\times$10$^{23}$\,J\,s$^{-1}$ & 30 & geometric \\ \hline
\end{tabular}
\end{table*}

\begin{figure}[ht!]
\centering
\includegraphics[width=\columnwidth]{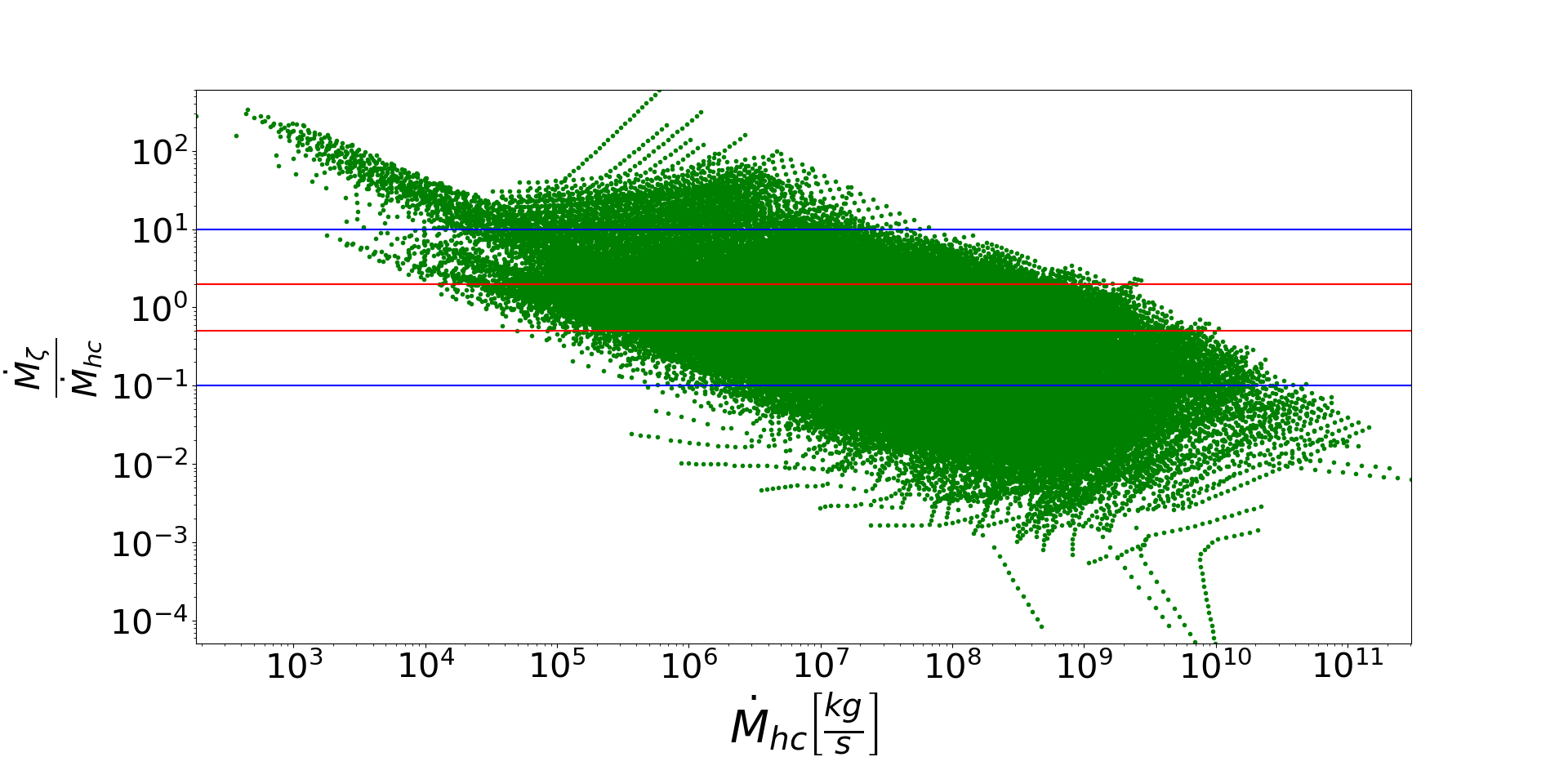}
\includegraphics[width=\columnwidth]{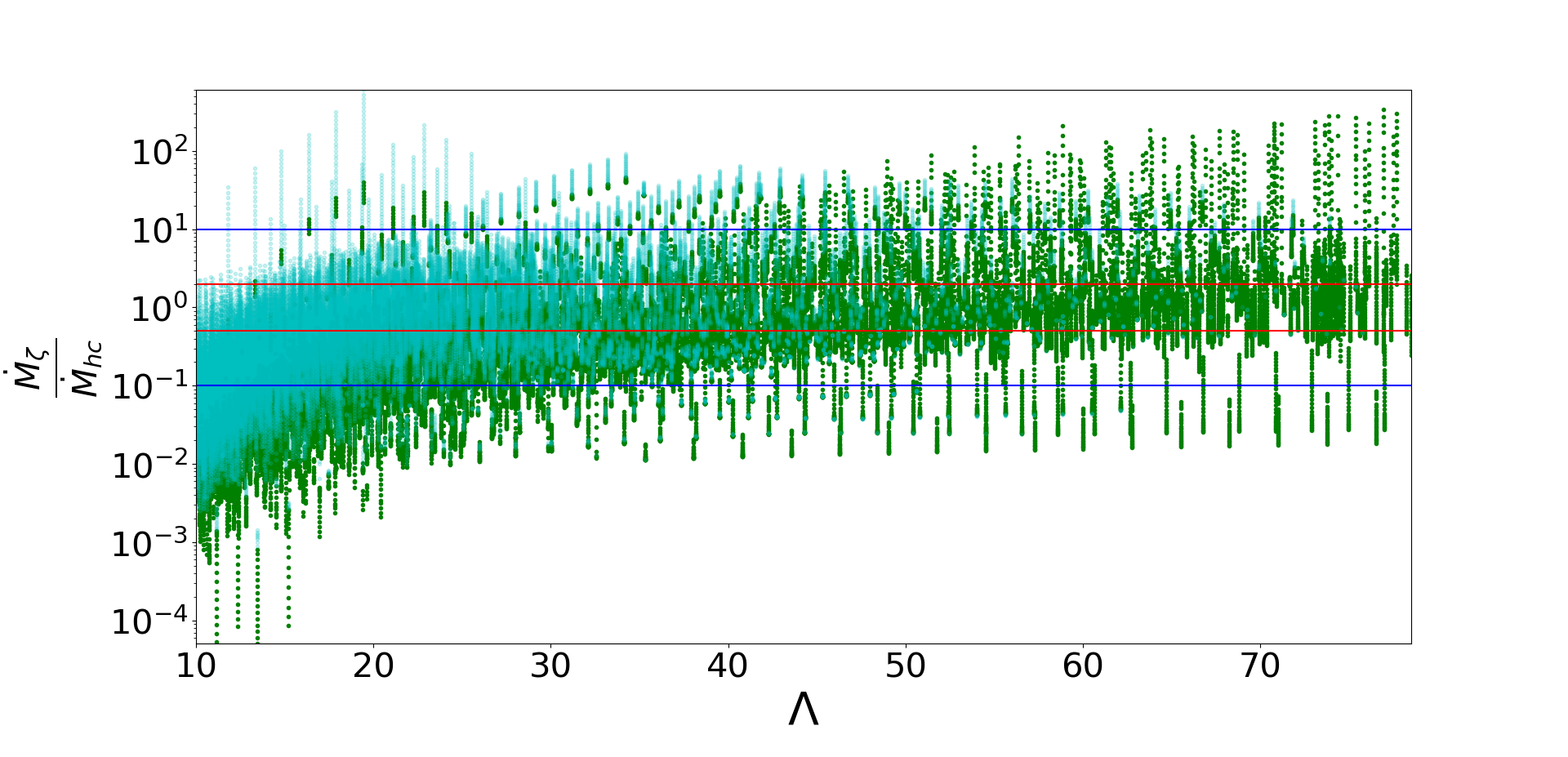}
\caption{Ratio of the mass-loss rates obtained following the energy-limited formalism ($\dot{M}_{\rm \zeta}$) and from hydrodynamic simulations ($\dot{M}_{\rm hc}$). Top: Plotted as a function of $\dot{M}_{\rm hc}$. Bottom: Plotted as a function of $\Lambda$. The dark green and bright blue points represent planets that do and do not comply with the subsonic condition, respectively (see Sect.~\ref{sec:inconsistencies}). In both panels, the red and blue lines mark the regions within which $\dot{M}_{ \rm \zeta}$ and $\dot{M}_{ \rm hc}$ differ by less than a factor of two and ten, respectively.}
\label{ratio_hc}
\end{figure}

The top panel of Fig.~\ref{ratio_hc} shows the $\dot{M}_{\rm \zeta}$-to-$\dot{M}_{\rm hc}$ mass-loss ratio as a function of the mass-loss rate obtained from interpolating across the grid of hydrodynamic models ($\dot{M}_{ \rm hc}$). In general, the mass-loss ratio ranges over several orders of magnitude and decreases with increasing $\dot{M}_{ \rm hc}$. The energy-limited formalism only reproduces the hydrodynamic results well for planets with moderate mass-loss rates, while it overestimates and underestimates escape by several orders of magnitude for weakly and strongly mass-losing planets, respectively. This is because for planets with low mass-loss rates the energy-limited formalism fails as a result of the hydrodynamic escape assumption being invalid. On the contrary, for planets with high mass-loss rates, the energy-limited formalism fails mostly as a consequence of boil-off taking over as the main escape-driving mechanism \citep[see also][]{fossati2017aeronomical,kubyshkina2018grid}. This is confirmed by the bottom panel of Fig.~\ref{ratio_hc}, which shows that for planets with low $\Lambda$ values, which are prone to boil-off escape, the energy-limited formalism severely underestimates mass-loss rates. 

\begin{figure}
\centering
\includegraphics[width=\columnwidth]{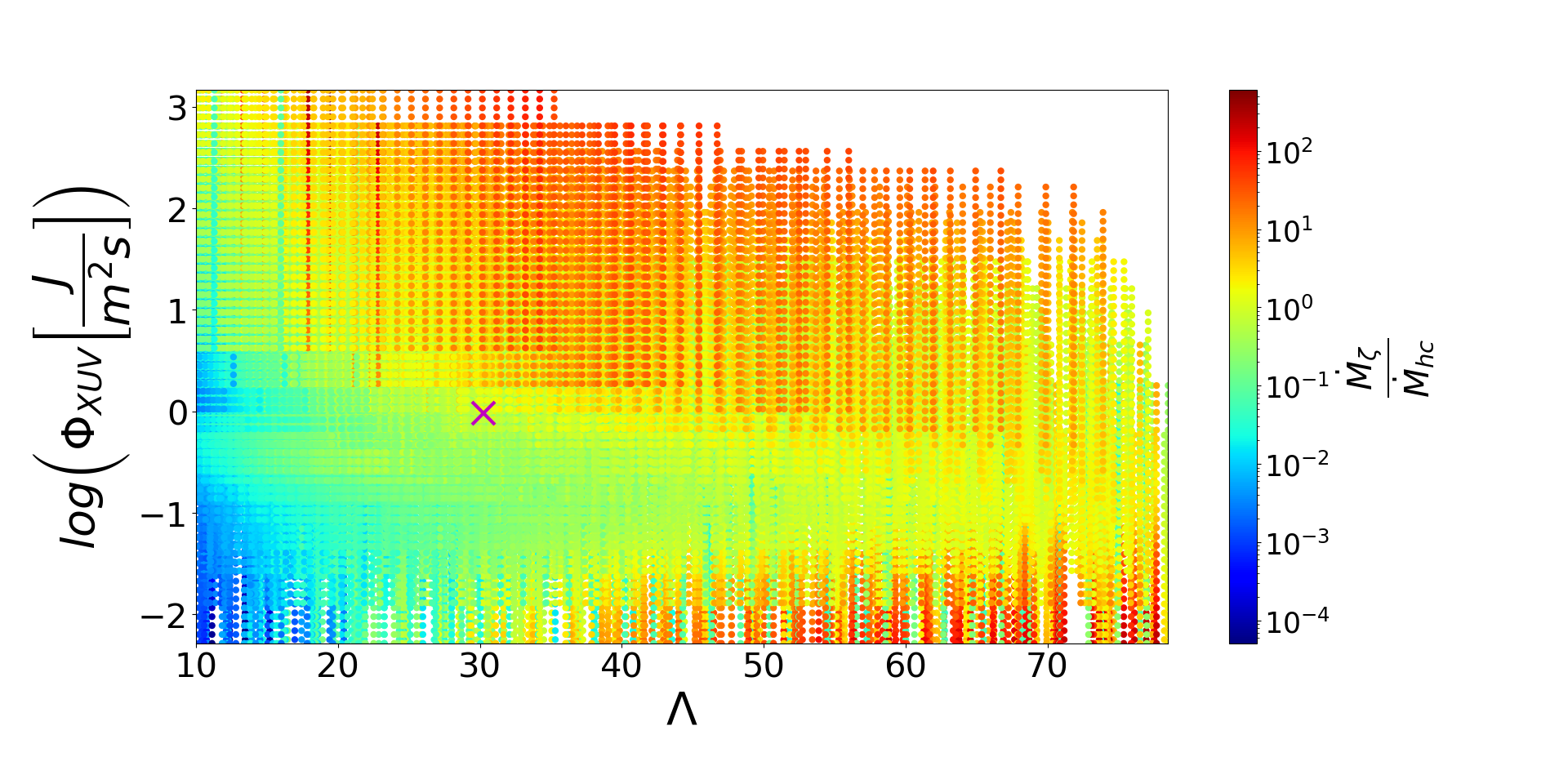}
\caption{Ratio of the mass-loss rates obtained following the energy-limited formalism ($\dot{M}_{\rm \zeta}$) and from hydrodynamic simulations ($\dot{M}_{\rm hc}$) as a function of $\Lambda$ and incident stellar XUV flux. For reference, the purple X shows the position of HD97658\,b \citep{kubyshkina2018overcoming}.}
\label{fig_ratio_phi_lambda0}
\end{figure}

Furthermore, Fig.~\ref{ratio_hc} shows that the energy-limited formalism also fails to reproduce the results of the hydrodynamic simulations for some planets with moderate mass-loss rates and intermediate $\Lambda$ values. We further explore this result in Fig.~\ref{fig_ratio_phi_lambda0}, which shows how $\dot{M}_{\rm \zeta}$/$\dot{M}_{ \rm hc}$ varies as a function of both $\Lambda$ and stellar XUV flux irradiation.

For planets with $\Lambda$ smaller than $\approx$30, the energy-limited approximation on average significantly underestimates the mass-loss rate. As mentioned above, this is mostly due to the fact that, for planets with low $\Lambda$ values, boil-off becomes the main escape-driving mechanism. In practice, the energy-limited approach assumes that the temperature profile presents a 0\,K minimum. This implies that any thermal energy already available in the atmosphere as a result of, for example, gravitational contraction and/or optical and infrared stellar irradiation is not taken into account. For planets characterised by low $\Lambda$ values, this thermal energy is significant with respect to the planetary gravitational energy, which drives boil-off and falsifies the assumption set by the energy-limited approach on the temperature profile. In contrast, the energy-limited approximation appears to perform better at low $\Lambda$ values and for high stellar XUV irradiation conditions. However, this stems from the fact that for such planets the energy-limited approach ignores boil-off and overestimates the upper atmospheric temperature, which have opposite effects on the mass-loss rate, finally returning mass-loss rates comparable to those given by the hydrodynamic simulations. We come back to this point in Sect.~\ref{sec_disc}.

\begin{figure}[ht!]
\centering
\includegraphics[width=\columnwidth]{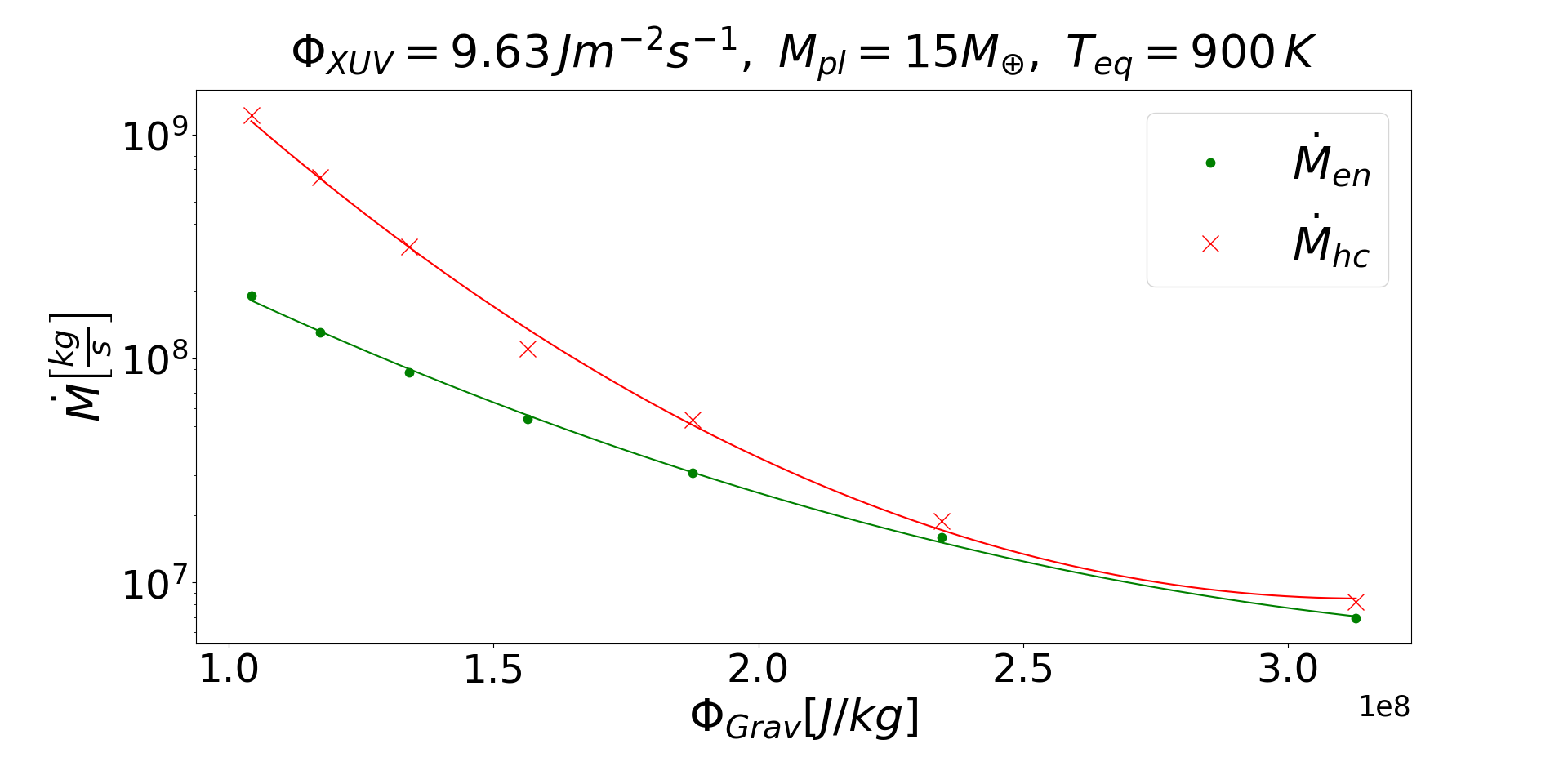}
\includegraphics[width=\columnwidth]{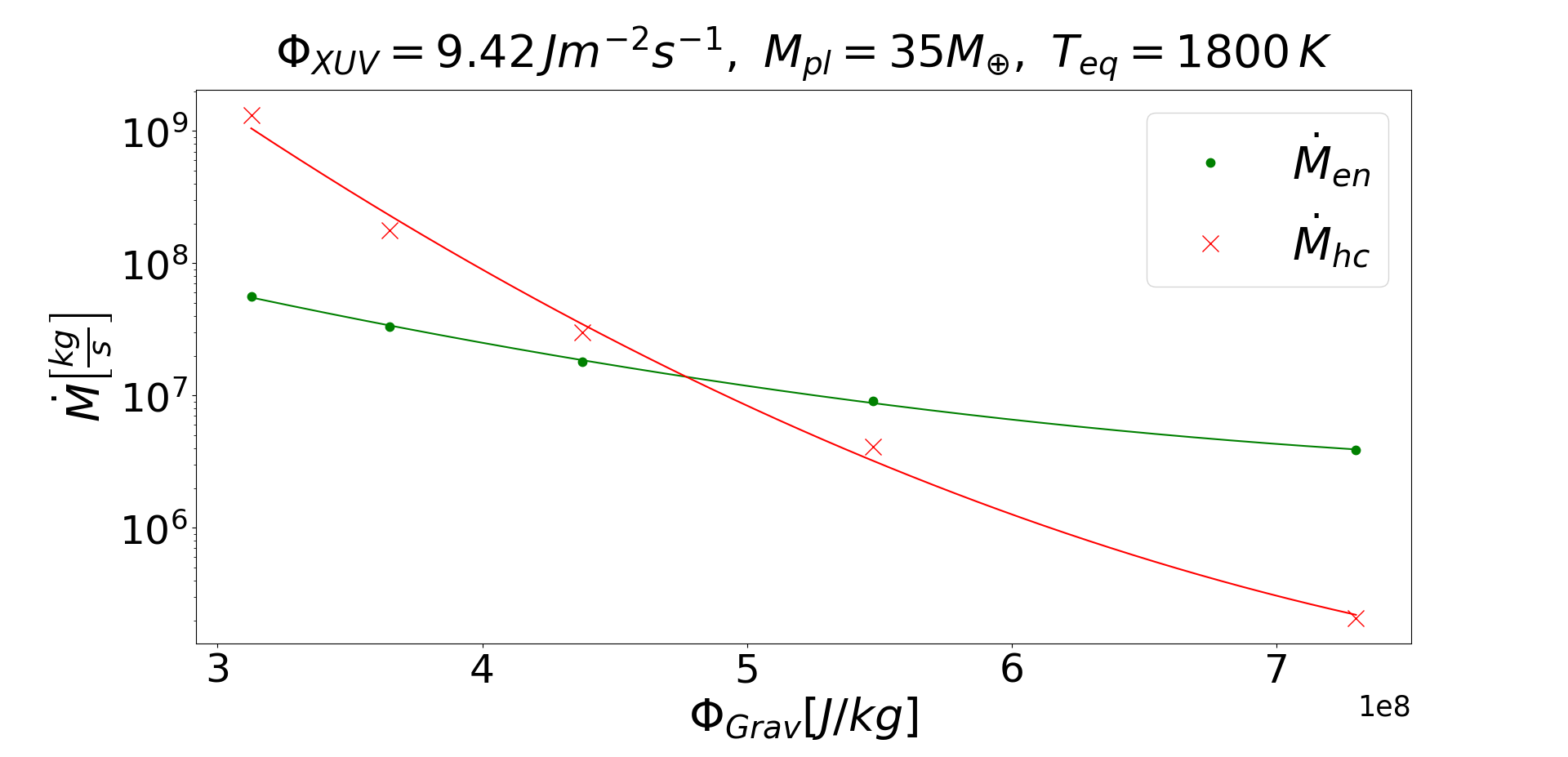}
\caption{Energy-limited (green dots) and hydrodynamic (red crosses) mass-loss rates as a function of gravitational potential (G$M_{\rm pl}$/$R_{\rm pl}$) for two samples of planets with the same mass, stellar XUV irradiation, and equilibrium temperature (the parameters are listed at the top of each panel). The green and red lines show quadratic fits to the energy-limited and hydrodynamic mass-loss rates, respectively, in the form $\log\dot{M}$\,=\,$p_0 + p_1\,\Phi_{\rm Grav} + p_2\,\Phi_{\rm Grav}^2$. The coefficients $p_0$, $p_1$, and $p_2$ of the fits are listed in Table~\ref{tb_fits}.}
\label{fig_gravpot}
\end{figure}

Figure~\ref{fig_ratio_phi_lambda0} shows that for planets with higher $\Lambda$ values, on average the energy-limited approximation reproduces the mass-loss rates given by hydrodynamic simulations within roughly an order of magnitude. This is the regime in which most of the assumptions involved in the energy-limited approach are valid. However, for a rather large number of planets characterised by low or high XUV flux irradiation levels, the energy-limited approximation appears to significantly overestimate mass loss, independently of $\Lambda$. This can be interpreted as follows. The main parameter setting the magnitude of the mass-loss rate of a given planet is the gravitational potential. Therefore, planets with different gravitational potentials, all else being equal, have significantly different mass-loss rates. However, $\dot{M}_{\rm \zeta}$ and $\dot{M}_{\rm hc}$ behave differently as a function of the planetary gravitational potential ($\Phi_{\rm Grav}$). This is shown in Fig.~\ref{fig_gravpot}, which presents, as an example, $\dot{M}_{\rm \zeta}$ and $\dot{M}_{\rm hc}$ as a function of $\Phi_{\rm Grav}$ for two sets of planets of the same mass, stellar XUV irradiation, and equilibrium temperature. In both cases, $\dot{M}_{\rm \zeta}$ and $\dot{M}_{\rm hc}$ behave differently as a function of $\Phi_{\rm Grav}$. At low $\Phi_{\rm Grav}$ values, $\dot{M}_{\rm hc}$ is consistently higher than $\dot{M}_{\rm \zeta}$ as a result of the fact that $\dot{M}_{\rm \zeta}$ does not account for boil-off. Instead, at high $\Phi_{\rm Grav}$ values, $\dot{M}_{\rm \zeta}$ is comparable to $\dot{M}_{\rm hc}$ in the case of lower mass planets and significantly larger than $\dot{M}_{\rm hc}$ in the case of higher mass planets. Furthermore, the relative behaviour of $\dot{M}_{\rm \zeta}$ and $\dot{M}_{\rm hc}$ varies as a function of the other system parameters. This explains why planets lying in the same place in the $\Lambda$-$\Phi_{\rm XUV}$ plane can present widely different $\dot{M}_{\rm \zeta}$/$\dot{M}_{\rm hc}$ values.

\begin{table}[ht!]
\centering
\caption{Coefficients of the quadratic fits shown in Fig.~\ref{fig_gravpot}.}
\label{tb_fits}
\begin{tabular}{c|cc}
\hline\hline   
     Coefficient & $\dot{M}_{\rm en}$ & $\dot{M}_{\rm hc}$ \\ 
     \hline
      & \multicolumn{2}{c}{Top panel}\\
     $p_0$ & 9.6 & 11.7 \\ 
     $p_1$ & $-$1.5$\times$10$^{-8}$ & $-$3.0$\times$10$^{-8}$ \\ 
     $p_2$ & 2.0$\times$10$^{-17}$ & 4.8$\times$10$^{-17}$ \\ \hline
      & \multicolumn{2}{c}{Bottom panel}\\
     $p_0$ & 9.4 & 1.4 \\ 
     $p_1$ & $-$6.5$\times$10$^{-9}$ & $-$2.0$\times$10$^{-8}$ \\ 
     $p_2$ & 3.5$\times$10$^{-18}$ & 1.0$\times$10$^{-17}$ \\ \hline
\end{tabular}
\end{table}

Therefore, the energy-limited escape approach is not suitable for describing planets with $\Lambda$ values smaller than about 30 because the assumptions involved in the approach lead to an inappropriate description of the actual thermospheric temperature profile. Instead, for planets with higher $\Lambda$ values, the energy-limited escape approach can be used to provide an upper limit on the mass-loss rates, but significant overestimation may be expected for some planets as a result of the improper dependence of the mass-loss rates on planetary gravitational potential. This indicates that the energy-limited formalism may not even be safe to use in cases where the main assumptions listed in Sect.~\ref{sec:assumptions} appear to be met; we analyse this fact in more detail in the next section.

\section{The potential use of the energy-limited approximation in estimating mass-loss rates} \label{sec_disc}
\subsection{Inconsistencies within the energy-limited escape approach}\label{sec:inconsistencies}
%
\begin{table*}[t]
\centering
\caption{System parameters and atmospheric physical conditions derived from the energy-limited approach of two planets with almost identical $\Lambda$ values and $\dot{M}_{\rm \zeta}$-to-$\dot{M}_{\rm hc}$ mass-loss ratios (close to 1). The heavier planet complies with the subsonic condition, namely the condition for which the XUV absorption radius lies below the sonic point, while the lighter planet does not.}
\label{tb_planets}
\begin{tabular}{c|c|c|c|c|c|c|c|c|c|c}
\hline\hline   
      Subsonic &  $M_{\rm pl}$ & $R_{\rm pl}$ & $T_{\rm eq}$ & $\Phi_{\rm XUV}$ & $\Lambda$ & $r_{\rm XUV}$ & $\dot{M}_{\zeta}$ & $\dot{M}_{\rm hc}$ & $T_{\rm XUV}$ & $T_{\rm limit}$ \\ 
      condition & [$M_{\rm \oplus}$] & [$R_{\rm \oplus}$] & [K] & [J\,m$^{-2}$\,s$^{-1}$] &  & [$R_{\rm pl}$] & [kg\,s$^{-1}$] & [kg\,s$^{-1}$] & [K] & [K] \\
      \hline
     Met & 19 & 5 & 1000 & 2.9 & 24.8 & 1.11 & 8.3$\times$10$^6$ & 8.7$\times$10$^6$ & 11,190 & 16,570 \\ \hline
     Not met & 13 & 4 & 900 & 39.5 & 24.5 & 1.03 & 7.0$\times$10$^7$ & 7.4$\times$10$^7$ & 10,683 & 115 \\ \hline
\end{tabular}
\end{table*}

\citet{Watson1981} employed a variety of assumptions to justify the simplifications that are necessary to derive the energy-limited approximation from the hydrodynamic equations. Therefore, it would appear that the energy-limited approximation can only be considered to reliably estimate mass-loss rates when the assumptions are met. However, this is not exactly the case. As a matter of fact, the energy-limited approximation sometimes appears to also reproduce the mass-loss rates provided by hydrodynamic simulations fairly well for planets that do not appear to comply with the energy-limited assumptions. For these cases, the match of the two mass-loss rates (i.e. $\dot{M}_{\rm \zeta}$ and $\dot{M}_{\rm hc}$) seems to occur by chance. In reality, this happens  either because the true planetary conditions comply with the energy-limited assumptions, though the results of the energy-limited equations indicate the opposite, or because wrong assumptions within the energy-limited approach cancel one another out. The problem lies in the fact that without hydrodynamic simulations it is not possible to tell whether the energy-limited mass-loss rates, computed for planets for which the energy-limited approach indicates that the assumptions are not met, are correct or not. This nullifies the need to use the energy-limited approach. This is why without hydrodynamic simulations it is nearly impossible to determine {\it a priori} whether certain planetary conditions would be such that the energy-limited approximation returns a mass-loss rate compatible, or not, with that given by hydrodynamic simulations or to determine why it is even more difficult to conclude whether a certain planet would comply with the assumptions of the energy-limited approach. We present below a few examples describing what is stated above. 

The subsonic condition, which assumes that the entirety of the stellar XUV energy is absorbed below the sonic point, is important within the energy-limited approach to justify neglecting the velocity term in the hydrodynamic equations. Within the energy-limited framework one can assess whether a planet complies with the subsonic assumption by using the following condition on the temperature at the XUV absorption radius ($T_{\rm XUV}$), which is derived (in SI units) following \citet{Watson1981} and Eq.~(\ref{eq_temperatureprofile}):
\begin{equation}
\label{eq_limit}
    T_{\rm XUV} < T_{\rm limit} = \frac{T_{\rm eq}^2}{1.2605 \ (\zeta \kappa_0)^2}\,.
\end{equation}
We remark that since this condition has been derived within the energy-limited framework, it is valid only as long as all the assumptions of the energy-limited approximation are met. This implies that it cannot be used to assess the applicability of the energy-limited approximation, but it can be used to check whether the energy-limited framework is consistent within itself once it is applied to a specific planet. 

The bottom panel of Fig.~\ref{ratio_hc} highlights in green planets that comply with the subsonic assumption and in light blue those that do not. Figure~\ref{ratio_hc} shows that there is a large number of planets in the grid to which the subsonic assumption does not apply, though their energy-limited mass-loss rates are in agreement with those given by hydrodynamic simulations. For these planets, the physical conditions at the XUV absorption height would imply a supersonic expansion of the gas, meaning that the velocity term in the hydrodynamic equations can no longer be neglected. As shown by Fig.~\ref{ratio_hc}, however, this inconsistency does not automatically imply that the energy-limited approximation fails to correctly estimate the mass-loss rates. 

We show this in practice by comparing, in Table~\ref{tb_planets}, two planets that have slightly different masses (19 and 13\,$M_{\oplus}$) but  similar values of $\Lambda$, similar $\dot{M}_{\rm \zeta}$-to-$\dot{M}_{\rm hc}$ mass-loss ratios (close to 1), similar equilibrium temperatures ($r_{\rm XUV}$), and similar $T_{\rm XUV}$ values. Despite the many similarities, the heavier planet complies with the subsonic assumption, while the lighter planet does not. Since the subsonic assumption appears to be violated for the lighter planet, one would  wonder why the energy-limited approximation returns a mass-loss rate in agreement with that of the hydrodynamic simulations. The reason is that the subsonic assumption is only violated within the energy-limited framework. In fact, because of the higher incident XUV flux of the lighter planet, the energy-limited approach artificially moves $r_{\rm XUV}$ closer to the planet, while in reality XUV absorption happens higher up in the atmosphere, where it is easier for the particles to escape, resulting in a more expanded atmosphere and similar mass-loss rates without the need to increase $T_{\rm XUV}$. What happens is that the combination of a larger $r_{\rm XUV}$ and a lower $T_{\rm XUV}$ allows for the velocity of the gas at $r_{\rm XUV}$ to remain subsonic, justifying neglecting the velocity term in the hydrodynamic equations. In this case, the first problem of the energy-limited approximation is not that it produces a significantly wrong mass-loss rate, but that it is not consistent within itself as it predicts a violation of its own assumptions. The second problem is that the user of the energy-limited approximation cannot realise this unless an independent, more reliable, and computationally expensive approach is used, cancelling out the benefits of using the energy-limited approximation.

Another assumption that does not seem to be met for a significant number of planets is the presence of a 0\,K minimum in the thermospheric temperature profile. \citet{Watson1981} argue that, although not realistic, such a temperature profile is the most efficient at maximising the amount of energy used to drive escape. As a matter of fact, this is the assumption that led them to the fact that the energy-limited approximation returns an upper limit on the XUV-driven mass-loss rate. A steeper temperature profile would result in negative temperatures, which are not physical, while a shallower profile would drain more of the absorbed XUV energy into conduction, heating the lower parts of the atmosphere. However, for example in the case of planets with very high mass-loss rates, the energy-limited approximation estimates that $r_{\rm XUV}$ lies very close to the planetary surface (middle panel of Fig.~\ref{fig_zeta_vs_rpl}) and that $T_{\rm XUV}$ is very high (top panel of Fig.~\ref{fig_temperature}). This results in extremely steep temperature profiles very close to the lower boundary, where the atmosphere is still very dense. Both conduction and convection in the collisionally dominated thermosphere make such a steep thermospheric temperature profile unlikely. 

The results of such an inadequate prediction of the temperature profile can, for example, be seen in the case of planets with large $\Lambda$ values that are subject to strong XUV irradiation. For these planets, the energy-limited approximation predicts that $r_{\rm XUV}$ lies close to the lower boundary (bottom panel of Fig.~\ref{fig_zeta_vs_rpl}) and that $T_{\rm XUV}$ reaches as high as 50,000\,K (top-right corner of the top panel of Fig.~\ref{fig_temperature}). The resulting steep temperature profile is very efficient at driving particle escape, leading to high mass-loss rates. However, when looking at the results of hydrodynamic simulations (Fig.~\ref{fig_ratio_phi_lambda0}), one can see that the mass-loss rate for these planets is on average overestimated by up to three orders of magnitude. In reality, what happens is that conduction and convection distribute the absorbed stellar energy across several atmospheric layers, leading to atmospheric expansion, which in turn results in a larger $r_{\rm XUV}$ and a shallower thermospheric temperature profile, and thus in lower $T_{\rm XUV}$ and more moderate mass-loss rates. For these planets, the energy-limited approximation fails to correctly predict the thermospheric temperature profile. Hydrodynamic simulations predict that the thermospheric temperatures of planets with low equilibrium temperatures remain below 1000\,K, while those of hot Jupiters do not exceed 20,000\,K \citep[e.g.][]{tian2005transonic,murray2009atmospheric,kubyshkina2018grid}. Therefore, there are cases for which the energy-limited approximation returns thermospheric temperatures that are significantly higher than what is predicted by more sophisticated models, which is caused by the fact that the energy-limited approximation is unable to correctly estimate temperature profiles and XUV absorption heights.

Finally, we briefly examine how well the energy-limited approximation estimates the location of $r_{\rm XUV}$ with respect to the atmospheric pressure, and thus optical depth to XUV radiation. The bottom panel of Fig.~\ref{fig_temperature} shows that, according to the energy-limited approximation, the position of the XUV absorption height lies at pressures ranging between 10$^{-9}$ and 10$^{-11}$\,bar. \citet{murray2009atmospheric} found that in a hydrogen atmosphere $P_{\rm XUV}$ can be estimated as (in SI units)
\begin{equation}
    P_{\rm XUV} = 5.3 \times 10^{16} \ m \ \frac{K \ G \ M_{\rm pl}}{R_{\rm pl}^2}\,,
\end{equation}
which typically results in pressures of about 10$^{-9}$\,bar, but they can also reach up to two orders of magnitude lower for planets with low equilibrium temperatures ($\sim$300\,K) and two orders of magnitude  higher for planets with high equilibrium temperatures. Therefore, there are planets within the considered grid for which the value of $P_{\rm XUV}$ estimated by the energy-limited approach is too low to be able to absorb the majority of the incoming XUV radiation.
\subsection{Ranges of applicability of the energy-limited escape approach}
\begin{figure}[ht!]
\centering
\includegraphics[width=\columnwidth]{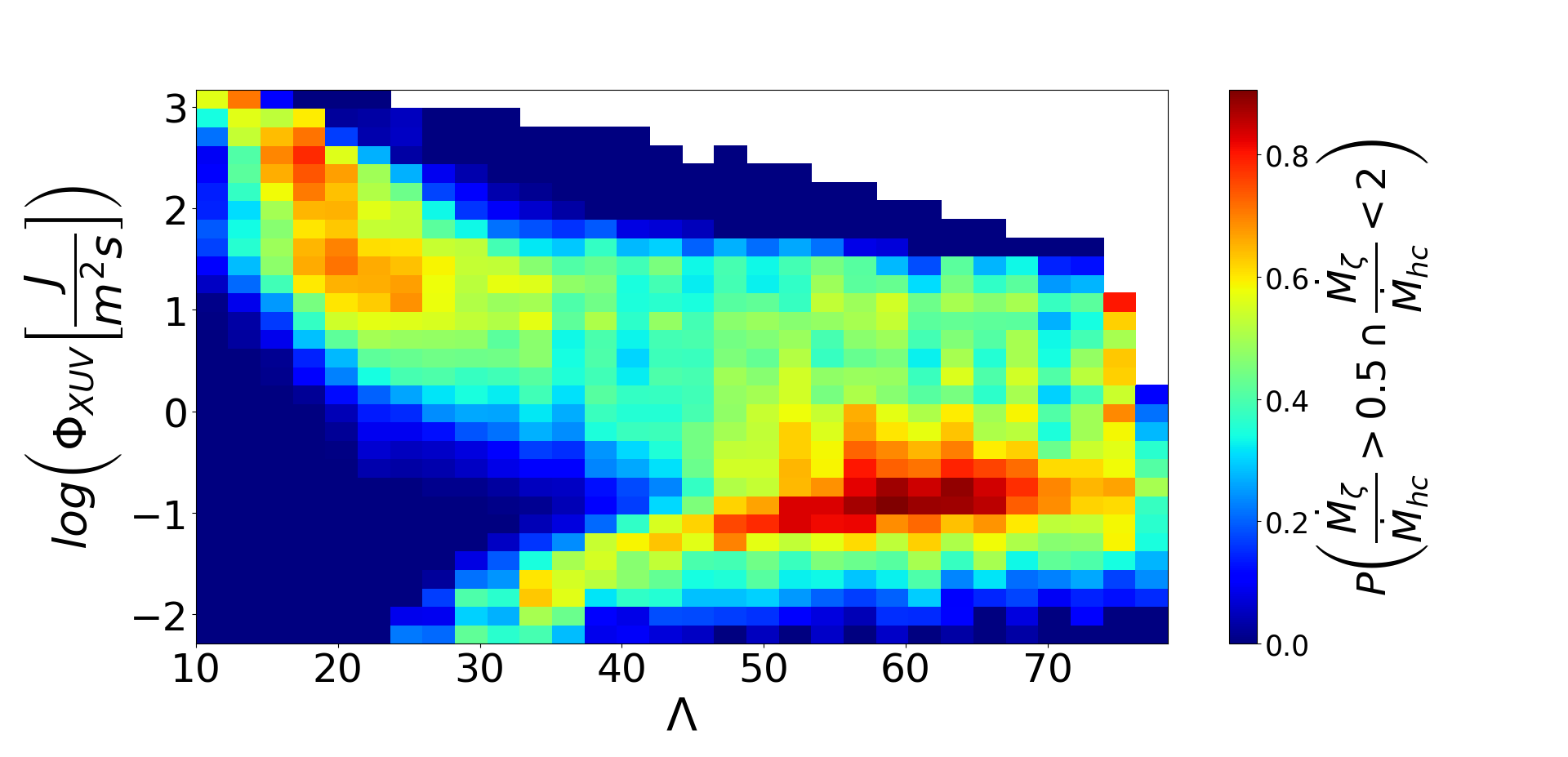}
\includegraphics[width=\columnwidth]{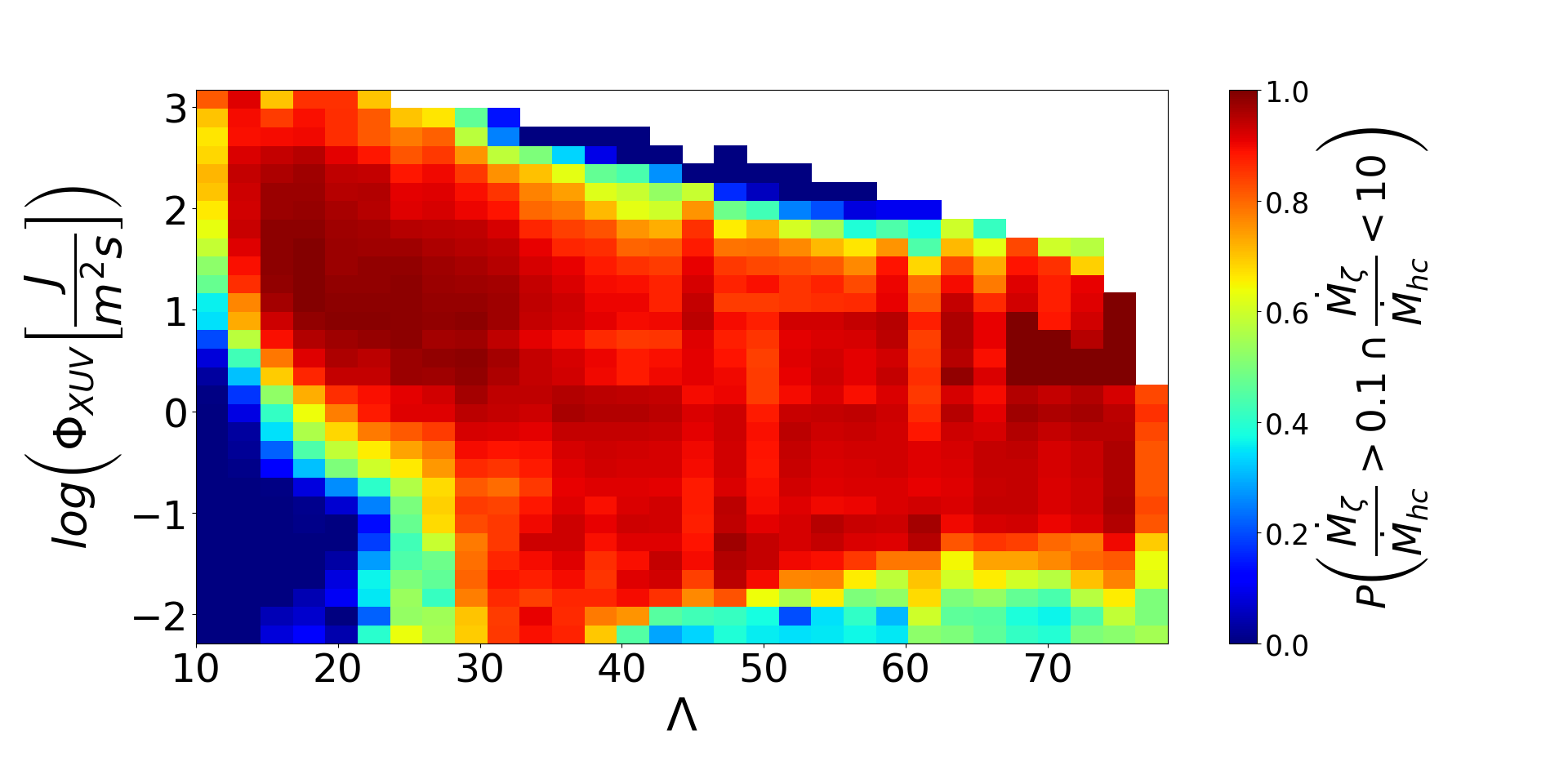}
\caption{Probability of $\dot{M}_{\zeta}$ agreeing with $\dot{M}_{\rm hc}$ within a certain factor as a function of $\Lambda$ and $\Phi_{\rm XUV}$. The probabilities are computed considering planets lying within 30 $\Lambda$ bins and 30 $\Phi_{\rm XUV}$ bins. Bins comprising fewer than ten planets have been discarded. Top: Agreement within a factor of two. Bottom: Agreement within a factor of ten.}
\label{fig_probability}
\end{figure}

To provide a more practical assessment of the parameter space in which the energy-limited approximation can be safely used to estimate atmospheric mass-loss rates, we computed the probability of obtaining $\dot{M}_{\zeta}$ values comparable to those obtained by hydrodynamic simulations for different combinations of planetary parameters within the grid defined by Table~\ref{tb_grid2}. Figure~\ref{fig_probability} shows the probability of $\dot{M}_{\rm \zeta}$ and $\dot{M}_{\rm hc}$ agreeing within a factor of two (top panel) or ten (bottom panel) as a function of $\Lambda$ and $\Phi_{\rm XUV}$. 

The top panel of Fig.~\ref{fig_probability} gives the impression that the energy-limited approximation can never be used to accurately predict mass-loss rates as there is essentially no region of the parameter space in which one can be sure of obtaining values within a factor of two of what is given by hydrodynamic simulations. Indeed, for the vast majority of the parameter space the probability of having $\dot{M}_{\zeta}$ and $\dot{M}_{\rm hc}$ agreeing within a factor of two is often lower than 50\%. However, we remark that the energy-limited approximation has not been developed to correctly estimate mass-loss rates, but rather only to provide reliable upper limits. Furthermore, \citet{Watson1981} developed the energy-limited approximation to study the case of Earth's atmospheric escape and could not foresee its use in a completely different context, such as that of exoplanets.

The bottom panel of Fig.~\ref{fig_probability} shows that the energy-limited approximation does a much better job when it is used to provide an order of magnitude estimate of the mass-loss rates. Indeed, there is a wide area within the considered parameter space in which $\dot{M}_{\zeta}$ and $\dot{M}_{\rm hc}$ agree within a factor of ten at least 75\% of the time. However, there does not seem to be a safe range of system parameters in which there is the certainty that $\dot{M}_{\zeta}$ and $\dot{M}_{\rm hc}$ match within a factor of ten.

\begin{figure}[ht!]
\centering
\includegraphics[width=\columnwidth]{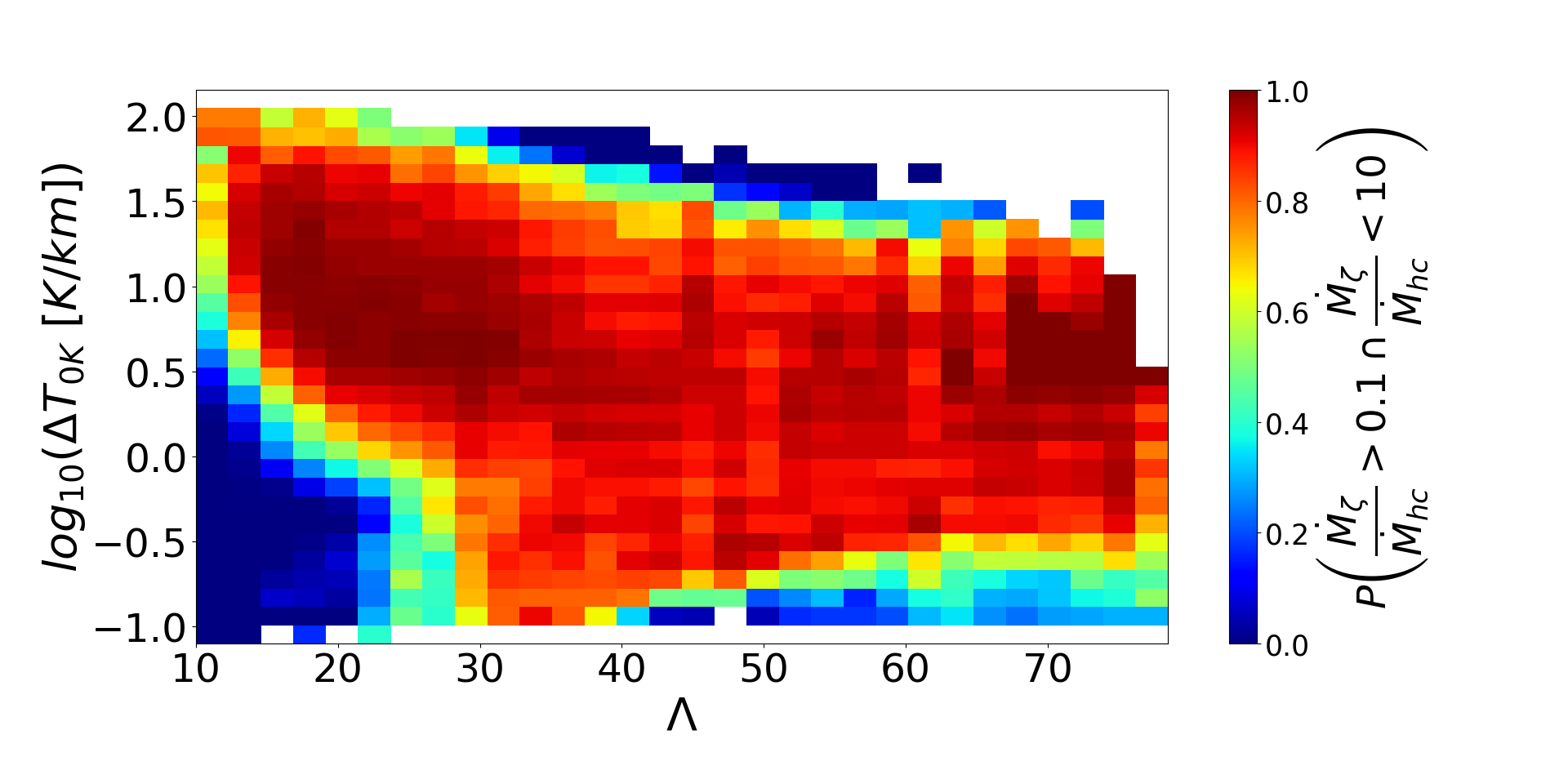}
\includegraphics[width=\columnwidth]{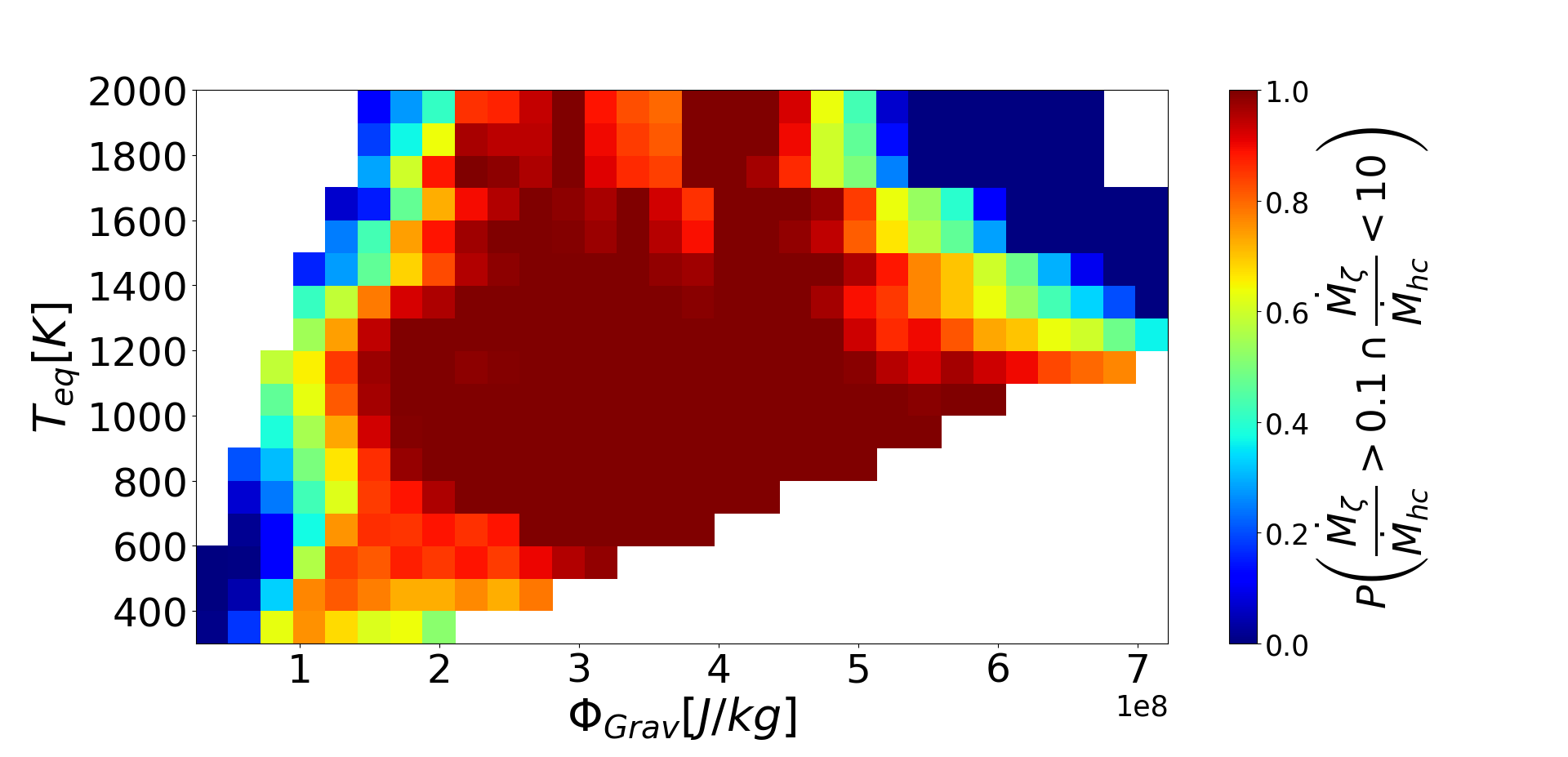}
\includegraphics[width=\columnwidth]{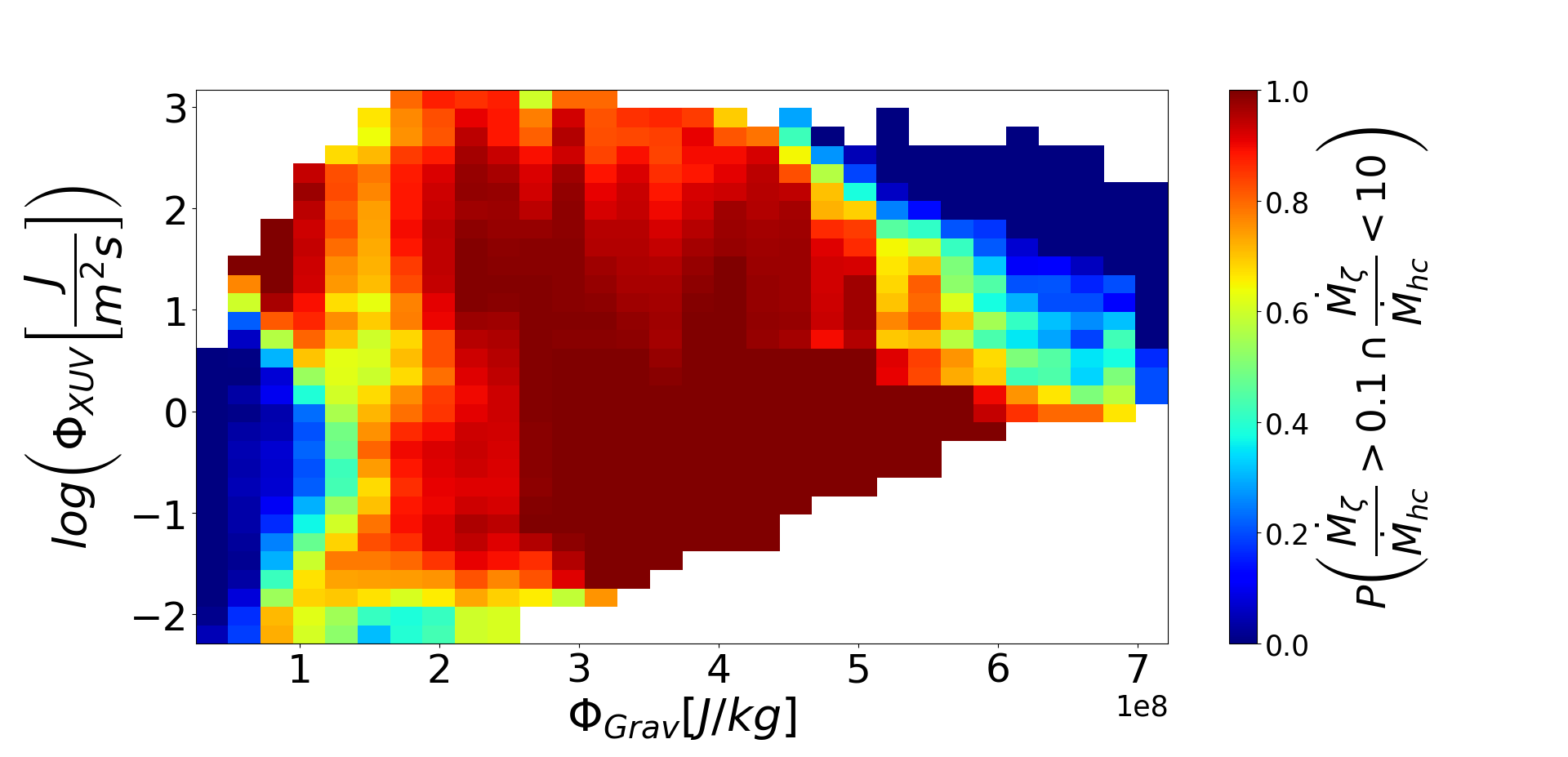}
\caption{Probability of $\dot{M}_{\zeta}$ being within a factor of ten of $\dot{M}_{\rm hc}$ as a function of different planetary parameters. Top: Plotted as a function of $\Lambda$ and $\Delta T_{0 K}$, i.e. the linear temperature gradient between the 0\,K minimum of the energy-limited temperature profile and the XUV absorption height. Middle: Plotted as a function of the gravitational potential of the planet corrected for tidal effects ($\Phi_{\rm grav}$) and $T_{\rm eq}$. Bottom: Plotted as a function of $\Phi_{\rm grav}$ and $\Phi_{\rm XUV}$. For all panels the probabilities are computed considering planets lying within 30 bins of each parameter, except for the equilibrium temperature. Here just 17 bins have been considered as only 18 distinct values of $T_{\rm eq}$ are represented in the planetary grid. Bins comprising fewer than ten planets have been discarded.}
\label{fig_tgrad}
\end{figure}

We assess here in greater detail the ranges of applicability of the energy-limited approximation. There are two distinct regions of the parameter space in which the energy-limited approximation appears to systematically provide incorrect mass-loss rates. The first one is located at the bottom-left corner of the bottom panel of Fig.~\ref{fig_probability} and is characterised by planets with low gravitational potential and low-to-intermediate XUV irradiation. The discrepancy in this regime between the energy-limited approximation and hydrodynamic simulations can be explained by looking at the resulting temperature profiles estimated within the energy-limited framework. We computed the linear temperature gradient from the predicted 0\,K minimum to the XUV absorption height (i.e. the maximum temperature) as a measure of the steepness of the corresponding temperature profile. The position of the 0\,K minimum is determined as \citep{Watson1981}
\begin{equation}
    \lambda_{0 K} = \frac{\beta}{\zeta \ \lambda_{XUV}^2}\,.
\end{equation}
The linear temperature gradient is then given by
\begin{equation}
    \Delta T_{0K} = \frac{T_{XUV}}{R_{XUV} - R_{0K}}\,.
\end{equation}
The top panel of Fig.~\ref{fig_tgrad} shows the probability of the energy-limited mass-loss rates being within a factor of ten of those given by hydrodynamic simulations as a function of $\Delta T_{0K}$ and $\Lambda$. Again, the bottom-left corner of this plot (low $\Lambda$ and low-to-intermediate $\Delta T_{0K}$ values) is populated by planets for which the energy-limited approximation returns mass-loss rates that are significantly different from those given by the hydrodynamic simulations. Therefore, we interpret this deviation as being related to the fact that the energy-limited approximation predicts too rapid a cooling of the lowest part of the atmosphere for planets that are not heated very strongly. As a matter of fact, the assumed 0\,K minimum of the energy-limited approximation leads to the amount of energy already present in the lower atmosphere not being used to drive escape, forcing the atmosphere to cool to 0\,K and then reheat up to the XUV absorption height. However, in reality, the lowest part of the atmosphere cools much more slowly and behaves nearly hydrostatically, leading to a more expanded atmosphere and an absorption of XUV energy farther from the lower boundary than predicted by the energy-limited approximation. The resulting temperature profile for planets with $\Lambda \approx$10 does not show a local temperature minimum but rather decreases monotonously from the lower boundary to the XUV absorption height. Furthermore, because of the low gravitational potential of these planets, the atmosphere already possesses almost enough energy at the lower boundary to drive the escape, and the additional energy from XUV absorption goes to increasing the already high mass-loss rates. Once the input energy (equilibrium temperature and XUV irradiation) decreases or the gravitational potential increases, the available energy is no longer sufficient to counteract adiabatic cooling at the lower boundary, leading to the appearance of a temperature minimum. The deeper the location of the minimum, the closer the mass-loss rate gets to what is predicted by the energy-limited formalism, generating a transition region in which the energy-limited estimates get increasingly more realistic with increasing $\Lambda$ and/or $\Phi_{XUV}$ values. Also, even for planets with low gravitational potentials, if $\Phi_{XUV}$ reaches sufficiently high levels, the energy-limited approximation predicts small XUV absorption radii, leading to steeper temperature profiles that are closer to reality and, thus, more realistic mass-loss rate estimates.

The second distinct range of parameters in which the energy-limited approximation consistently fails in reproducing the results of hydrodynamic simulations is covered by planets characterised by high gravitational potential and high XUV irradiation. The middle and bottom panels of Fig.~\ref{fig_tgrad} show the probability of the energy-limited mass-loss rates being within a factor of ten of those given by hydrodynamic simulations as a function of $\Phi_{\rm grav}$, $T_{eq}$, and $\Phi_{XUV}$, with the last two parameters being tightly related through the orbital separation. Planets with high gravitational potential and high irradiation are located in the top-right corner of each panel. We remark that the combination of a high gravitational potential and high equilibrium temperature still results in moderate $\Lambda$ values, which is why Fig.~\ref{fig_probability} does not highlight this part of the parameter space. These planets absorb a large amount of energy at the XUV absorption height, but their atmospheres are rather dense, because of the high gravitational potential, leading to an absorption height lying close to the lower boundary. The energy-limited approximation still assumes the presence of a 0\,K minimum, but a significant amount of XUV radiation is then absorbed at a short distance from the temperature minimum, and most of this energy must be used to drive escape. If only a small amount of the absorbed energy is used to drive the escape, the atmosphere will conduct more energy towards the lower boundary, preventing the formation of the 0\,K minimum. Therefore, the restriction imposed by the energy-limited approximation to downward conduction leads to unphysically high temperatures at the XUV absorption height (up to 50000\,K) and thus to excessively large mass-loss rates. In reality, thermal conduction is significant, particularly for these planets, and planetary atmospheres never reach a 0\,K minimum, leading to lower temperatures at the XUV absorption height, shallower temperature profiles, and less energy used to drive escape, finally resulting in mass-loss rates lower than those predicted by the energy-limited approximation.

We conclude that in the case of low-to-intermediate equilibrium temperatures and incident XUV fluxes, the energy-limited formalism does not provide reliable mass-loss rates for planets characterised by excessively high or low gravitational potential. We find that the main reason for which the energy-limited mass-loss rates deviate most strongly from those given by hydrodynamic simulations is the inadequate representation of the temperature profile, mostly driven by the assumption of the presence of a 0\,K minimum. Outside of those regions of the parameter space, the energy-limited mass-loss rates reproduce those given by hydrodynamic simulations fairly well in most cases. However, we remark that the boundaries within which the results of the energy-limited formalism are comparable to those of hydrodynamic simulations are not particularly sharp, and that there are extended transition regions in which energy-limited mass-loss rates are unreliable.

%
\section{Conclusion}\label{sec_concl}
We reviewed the energy-limited atmospheric escape approach \citep{Watson1981,erkaev2007roche}, which is widely used for estimating mass-loss rates for a broad range of planets that host hydrogen-dominated atmospheres, as well as for performing atmospheric evolution calculations. In particular, we listed and thoroughly described the assumptions employed by \citet{Watson1981} to simplify the hydrodynamic equations in the energy-limited formalism. We also studied the consequences of further approximating the energy-limited mass-loss formula by substituting the XUV absorption radius with the planetary radius. We found that this additional approximation leads to an underestimate of the energy-limited mass-loss rates by a factor of 25 at most. This result was obtained by considering planets that cover a very wide parameter space in terms of planetary mass, radius, and equilibrium temperature, the values of which range between 0.5\,$M_{\oplus}$ and 1\,$M_{\rm \jupiter}$, 1\,$R_{\oplus}$ and 1\,$R_{\rm \jupiter}$, and 300 and 2000\,K, respectively.

Finally, we considered a smaller grid of planets (planetary mass, radius, and equilibrium temperature ranging between 1\,$M_{\oplus}$ and 39\,$M_{\oplus}$, 1\,$R_{\oplus}$ and 10\,$R_{\oplus}$, and 300 and 2000\,K, respectively) to compare in detail the results of the energy-limited approximation with those of hydrodynamic simulations. For this comparison we considered the grid of models computed by \citet{kubyshkina2018grid}, who also provided an interpolation routine that enables the extraction of key simulation results for any planet lying within the grid boundaries.

This comparison led us to conclude that the energy-limited approximation can be used as an order of magnitude estimate of the mass-loss rate for planets with intermediate gravitational potentials and low-to-intermediate equilibrium temperatures and XUV irradiation fluxes. The energy-limited formalism fails to approximate mass-loss rates for planets outside of this parameter space because for these planets it wrongly predicts the atmospheric temperature profile as a result of forcing the presence of a 0\,K temperature minimum in the thermosphere.

This result allows the energy-limited approximation to still be used for single planets, as long as their planetary parameters are within the parameter space of reliable estimation presented here. However, it precludes the use of the energy-limited approximation for planetary evolution calculations that require computing mass-loss rates for planets covering a wide parameter space; it is very likely that the approximation is going to fail in some of those cases, leading to significantly inaccurate planetary atmospheric evolutionary tracks \citep[e.g.][]{kubyshkina2020}. Therefore, in order to reliably estimate mass-loss rates over a broad range of parameters, we recommend using different tools, such as hydrodynamic models, or approximations obtained on the basis of hydrodynamic simulations, such as those presented by \citet{kubyshkina2018grid} or \citet{kubyshkina2018overcoming}, which, being analytical, can be incorporated into planetary evolution models, as done for example by \citet{kubyshkina2018grid}, \citet{kubyshkina2019a}, \citet{kubyshkina2019b}, \citet{kubyshkina2020}, and \citet{modirrousta2020a,modirrousta2020b}.


\begin{acknowledgements}
DK acknowledges funding received from the European Research Council (ERC) under the European Union’s Horizon 2020 research and innovation programme (grant agreement No 817540, ASTROFLOW).
\end{acknowledgements}


\bibliographystyle{aa} 
\bibliography{references} 

\end{document}